\documentclass{aa}  
\usepackage{multirow}
\usepackage[dvipsnames]{xcolor}
\usepackage{graphicx}
\usepackage{natbib}

\usepackage{txfonts}
\usepackage{placeins}

\usepackage{subcaption}

\usepackage{hyperref}
\hypersetup{
  colorlinks,
  filecolor=blue,
  citecolor=blue,
  linkcolor=blue,
 urlcolor=blue}

\begin{document}

   \title{Negative gas-phase metallicity gradients in the narrow line region and galactic disc of local AGN-host galaxies}
   \titlerunning{Negative gas-phase metallicity gradients in local AGN-host galaxies}

\author{A.~Amiri\inst{1}    
\inst{,} \inst{2}
  \thanks{\emph{ Email:}
     amirnezamamiri@gmail.com}
  \and J.~H.~Knapen \inst{2} \inst{,} \inst{3}
 \and B.~D.~Lehmer \inst{1}
 \and A.~Khoram \inst{4} \inst{,} \inst{5}
}
\authorrunning{Amiri et al.}

\institute{Department of Physics, University of Arkansas, 226 Physics Building, 825 West Dickson Street, Fayetteville, AR 72701, USA
 \and Instituto de Astrofísica de Canarias E-38205, La Laguna, Tenerife, Spain
 \and Departamento de Astrofísica, Universidad de La Laguna, E-38200, La Laguna, Tenerife, Spain
 \and Department of Physics \& Astronomy "Augusto Righi", University of Bologna, via Gobetti 93/2, 40129 Bologna, Italy
 \and INAF, Astrophysics and Space Science Observatory Bologna, Via P. Gobetti 93/3, I-40129 Bologna, Italy}

% \abstract{}{}{}{}{} 
% 5 {} token are mandatory
\abstract{The gas-phase metallicity distribution in galaxies provides significant information on their evolution. We report the discovery of negative radial gradients in the gas-phase metallicity of the narrow-line region of the nine galaxies in the Measuring Active Galactic Nuclei Under MUSE Microscope (MAGNUM) galaxies: Centaurus A, Circinus, IC~5063, NGC~ 1068, NGC~1365, NGC~1386, NGC~2992, NGC~4945, NGC~5643. From strong-line abundance relations for active galactic nuclei (AGN) and star-forming regions, along with emission-line ratio diagnostics, we determine spatially resolved gas-phase metallicities for the kinematic components, galaxy disc and outflow. These relations involve sensitive strong emission lines, specifically [O\,{\sc iii}]$\lambda$5007, [N\,{\sc ii}]$\lambda$6584, H$\alpha$, H$\beta$, [S\,{\sc ii}]$\lambda$6716, and [S\,{\sc ii}]$\lambda$6731. The existence of  predominantly negative radial metallicity gradients in these AGN host galaxies indicates that metals are not necessarily moved from the central regions to the outskirts by AGN activity and that the gas-phase metallicity in galaxies may follow the general inside-out star formation scenario.}


   \maketitle
%
%-------------------------------------------------------------------

\section{Introduction}
The gas-phase metallicity ($Z_{\rm gas}$) is a key tracer of chemical evolution in galaxies, providing crucial insights into the processes that regulate star formation, gas inflow/outflow, and feedback mechanisms in galaxies \citep[e.g.][]{Mayor_1976,Mayor_vigrox_1981,Pagel_1979,Shaver_1983,Moustakas_2011_metallicity,cresci_z,Sommariva_2012_stellar_to_gas_z,khoram2025_1}. The radial gradient of $Z_{\rm gas}$ provides critical information about the evolutionary history of the galaxy metal content. Typically, spiral galaxies exhibit a negative metallicity gradient, with higher metal content in the central regions and lower metallicity at larger radii (i.e., accretion begins in the centre of a galaxy and expands outwards \cite{Matteucci}. This trend is often interpreted in the context of an inside-out galaxy scenario \citep[e.g.][]{Samland_1997,Portinari_1999,Boissier_2000,Pilkington_2012,bible_of_metallicity,Franchetto_2020}, where the central regions undergo more prolonged star formation and experience continuous metal enrichment. In this case, metals are produced in galaxies and returned to the inter stellar medium (ISM) through a variety of mechanisms, such as stellar winds from massive stars \citep[e.g.][]{Higgins_2023},supernova explosions \citep[]{Hillebrandt_supernova,Woosley_supernova}, neutron star mergers \citep{Thielemann_nutron_merging}, and the ejection of gas by asymptotic giant branch stars \citep[e.g.][]{Winckel_AGB}. 

Negative metallicity gradients, a common feature in the disks of most spiral galaxies, have been extensively studied \citep{Kennicutt_2003}. Notably, three major surveys, namely CALIFA \citep[Calar Alto Legacy Integral Field Area][]{Sanchez_2012}, MaNGA \citep[Mapping nearby Galaxies at Apache Point Observatory;][]{Bundy_2015}, and SAMI \citep[Sydney-AAO Multi-Object Integral-field spectrograph;][]{Bryant_2015}, have substantially expanded the dataset for measuring metallicity gradients in the local Universe. The primarily results from these surveys indicate that metallicity gradients predominantly exhibit a negative trend in nearby galaxies \citep{Mingozzi_2020,Belfiore_2021,2025Khoram_Belfiore}.Some studies report a distinct positive metallicity gradient, in contrast to others \citep{Cresci_2010, Queyrel_2012,Belfiore_2017,Carton_2018,Wang_2019,Mingozzi_2020,Simons_2021}. 

According to the scenario proposed by \cite{Somerville}, galaxies evolve within a dense environment where gas circulates through, around, and within them. Each phase in this cycle, such as altered star formation, accretion processes, and mergers, exerts a unique influence on the galaxy evolution. The presence of positive metallicity gradients may indicate direct accretion of metal-poor gas into the central regions of galaxies, resulting in diluted central metallicities and a pronounced change in gradients at small galactocentric distances \citep{Simons_2021}. \cite{Storchi_metal_poor_bh_feed} demonstrated that the interaction with gas-rich dwarf galaxies can initiate nuclear activity in host galaxies, leading to the accretion of low-metallicity gas into the nuclear regions and leading to its dilution. Some studies of high-redshift star-forming (SF) galaxies, \citep[e.g.][]{Queyrel_2012,Ju_2025,Sun_2025}, observed positive metallicity gradients and suggested that environmental interactions may induce an inversion in metallicity gradients.
$Z_{\rm gas}$ gradients may also have a flat profile. Both active and non-active spiral galaxies can exhibit flat metallicity gradients in their disks, where the metallicity remains nearly constant across the radial extent of the galaxy \citep{Nascimento_manga,Venturi_2024}. Such a distribution may indicate efficient metal mixing, likely driven by turbulence induced by AGN feedback \cite[e.g.][]{He_2019,He_2022,Ubler_2023}, galactic-scale outflows \citep[e.g.][]{Gibson_2013}, or environmental interactions \citep{Rich_JA2012,Queyrel_2012,Tores_2013}. 
Additionally, galaxies in dense environments, such as galaxy clusters, may experience enhanced mixing due to tidal interactions and ram--pressure stripping, further contributing to metallicity distribution \citep{Franchetto_2021a,Khoram_2024}.

In AGN host galaxies, metallicity investigations help distinguish between the effects of AGN-driven feedback and star formation on the enrichment of the ISM. AGN play a fundamental role in galaxy evolution by influencing gas dynamics through radiation pressure, outflows, and jets, which in turn impact the metal distribution within galaxies \citep[e.g.][]{Sanchez_2014,Choi_2020,Villar_2024,Amiri_2024,Marconcini_2025}. Understanding how $Z_{\rm gas}$ is distributed in AGN hosts is essential for constraining feedback mechanisms and their roles in modulating star formation and chemical enrichment. The enhanced metal enrichment attributed to AGN activity might result from mechanisms like an in-situ top-heavy initial mass function (IMF) in the accretion disk around the supermassive black hole \citep{Nayakshin_2005} or dust destruction in the broad line region (BLR), releasing metals into the ISM \citep{bible_of_metallicity}. This suggests that AGN can trigger rapid star formation and enrich the ISM.

Optical emission line studies have frequently revealed the presence of ionized gas outflows, particularly in the narrow line regions (NLRs) and extended NLRs, providing vital insights into the AGN-host galaxy correlation \citep{Veilleux,unger_1987,Pogge1988}. Resolved observations have enabled the detailed resolution of NLRs in nearby galaxies within close proximity to the supermassive black hole (SMBH), unravelling their extension over several kiloparsecs (kpc) into the bulges and/or disks of host galaxies \citep{Kang_2018}. The kinematic behavior and physical properties of  the ionized gas within the NLR yield crucial information on outflow characteristics, including the associated energy budgets \citep{Bennert_2006,Vaona,Dopita_2002,Mingozzi_2019,Chen_2019,Zhang_2022,Meena_2023}. AGN-driven outflows ejecting high-metallicity gas from the BLR at kpc scales can further enrich the NLR, potentially through in-situ star formation occurring within AGN-driven outflows \citep{Maiolino_2017,Gallagher_2019,Amiri_2024}.\\
The primary purpose of this work is to determine whether AGN-driven outflows significantly impact $Z_{\rm gas}$ gradients, by directly comparing the gradients in the discs and outflows of nearby Seyfert galaxies. Our main goal is to compare $Z_{\rm gas}$ gradients in the discs and outflow components (redshifted and blueshifted) of nine AGN host galaxies, in order to test whether AGN-driven outflows redistribute metals on galactic scales or whether the gradients remain similar to those in non-active spirals. This study also provides a modification over the analysis of \cite{Nascimento_manga}, who measured metallicity gradients in MaNGA galaxies without explicitly accounting for the influence of outflows. The present study expands our previous one of an individual AGN \citep{Amiri_2024} by examining a more extensive, homogeneous sample from the MAGNUM survey and by directly distinguishing between disc and outflow kinematic components through the emission-line profile decomposition method proposed by \cite{Mingozzi_2019}.

The paper is organized as follows: in Section \ref{sec:obs}, we provide a brief overview of the AGN  sample which is used in our study. In Section \ref{sec:obs_classification}, we outline how we classify the data into AGN and SF regions. In Section \ref{sec:results_II}, we further detail the calibration relations to estimate the gas-phase metallicities. In Section \ref{radial_Z}, we demonstrate how gas-phase metallicities depend on the radial distance. Finally, in Section \ref{sec:conclusion}, we discuss our findings and summarize the implications and significance of our study.
 
%--------------------------------------------------------------------
\section{Observational data}
\label{sec:obs}
We examine the ISM characteristics of nine nearby Seyfert galaxies, all of which have noticeable outflows by \cite{Mingozzi_2019}, with the Measuring Active Galactic Nuclei Under MUSE Microscope (MAGNUM) survey. To select those galaxies,  \cite{Mingozzi_2019} cross-referenced the Swift-BAT 70-month Hard X-ray Survey \citep{Baumgartner_2013}  with optically selected AGN samples from \cite{Maiolino_Riek_e1995} and \cite{Risaliti_1999}. The selection process focused on sources that could be observed from the Paranal Observatory and had a luminosity distance within 50 Mpc. Table\ref{table1} provides each galaxy's primary features, and the following gives a brief description of each galaxy.

\textbf{Centaurus A}: Located at a distance 3.82 Mpc, Centaurus A is one of the closest Seyfert 2 galaxies. The MAGNUM survey detected ionized outflows with complex kinematics \citep{Mingozzi_2019}, indicative of active interactions between the AGN and the surrounding ISM. Studies \citep[e.g.][]{Santoro_2015,Santoro_2016,Mingozzi_2019} have explored the host galaxy nuclear region, revealing intricate structures and dynamics associated with its active nucleus.

\textbf{Circinus}: A nearby gas rich Seyfert 2 galaxy at a distance of 4.2 Mpc with a spiral morphology. \cite{Matt_2000} evidenced an obscured AGN via X-ray spectral analysis. A study by \cite{Kakkad_2023_circinus} demonstrated the morphological features of ionized gas outflows in the Circinus galaxy using the MUSE in the Narrow Field Mode. Their analysis of the MUSE observations reveals a distinct collimated and clumpy outflow structure emerging near the nuclear region and extending up to approximately 30 pc in the northwest direction. This collimated configuration then transforms into two filamentary structures, resulting in an overall tuning-fork morphology of the outflow gas. \cite{Kakkad_2023_circinus} suggested that the collimated structure itself may result from interactions between jets and the ISM on small scales.

\textbf{IC~5063}: This early type galaxy, approximately 45.3 Mpc distant, exhibits Seyfert 2 characteristics. Since the radio plasma jet is expanding into a clumpy gaseous medium, it may be creating a cocoon of shocked gas that is pushed away from the jet axis, making this galaxy one of the most spectacular examples of jet-driven outflow, with similar features in the ionized, neutral atomic, and molecular phase \citep{Morganti_1998,Sharp_2010,Tadhunter_2014,Oosterloo_2017,Venturi_2021}. Using JWST MIRI data of the nuclear region of IC~5063, \cite{Dasyra_2024} demonstrated that an unidentified source at the galaxy nucleus emits a bright continuum that can be attributed to the torus with a possible contribution from the radio core.

\textbf{NGC~1068}: A barred spiral galaxy located approximately at 12.5 Mpc, NGC~1068 is one of the most studied Seyfert galaxies due to its proximity and brightness. Recent studies have revealed that low-power radio jets in NGC~ 1068 can drive significant turbulence in the ISM, leading to enhanced line widths perpendicular to the jets and indicating strong jet-ISM interactions \citep{Venturi_2021,Melso_2024}.

\textbf{NGC~1365}: Known as the great barred spiral galaxy, Seyfert 1.8 NGC 1365 is located at 18.6 Mpc distance. 
The MAGNUM survey observed ionized gas outflows with complex kinematics, suggesting interactions between the AGN activity and the host galaxy ISM. The MUSE data of this source reveal a bi-conical AGN-ionized outflow extended on several kpc as well non-circular motions compatible with inflows along the galactic bar \citep{Venturi_2018,Schinnerer_2023}.

\textbf{NGC~1386}: This spiral galaxy, approximately 15.6 Mpc away, hosts a Seyfert 2 nucleus. At a scale of a few parsecs from the center, \cite{Ardila_2017} made a  the direct measurement of powerful mass outflows traced by the forbidden high-ionization gas. They discovered two symmetrical expanding hot gas shells traveling in opposing directions along the line of sight \citep{Mingozzi_2019}. The gas shells' spatial correlation with X-rays and radio emission observed at the same parsec-scales indicates that this is a shock-driven outflow that was most likely caused by a newly formed core jet. 

\textbf{NGC~2992}: With a spiral morphology at a distance 31.5 Mpc, this galaxy contains a Seyfert 1.9 nucleus. H$\alpha$ and O III emission and soft X-ray observations \citep[e.g.][]{Colina_1987,Colbert_1996,Veilleux_2001} provide evidence of a broad biconical large-scale outflow that extends above and below the galaxy plane and which may be caused by AGN activity \citep{Mingozzi_2019,Friedrich_2010}. \cite{Luminari_2023ApJ}  suggested that the wind might drive feedback impacts between the AGN and the host galaxy by utilizing the combination of XMM-Newton and NuSTAR data. 

\textbf{NGC~4945}: This Seyfert 2 spiral galaxy  located at 3.7 Mpc is well known for being one of the nearest objects where starburst and AGN activity overlap. The absence of UV photons relative to X-rays and the fact that the AGN  presence is only confirmed by X-ray observations \citep[e.g.][]{Guainazzi_2000} suggest either unusual activity or the complete obscured of its UV emission along all lines of sight \citep{Marconi_2000}. Near- and mid-infrared spectroscopy, \citep{perez_2011} reveals a prominent dust lane aligned along the major axis of the galactic disc, which is linked to the region's very strong obscuration in the center. The biconical outflow that characterizes this galaxy is evident from the [NII] emission map \citep{Venturi_2017,Mingozzi_2019,Marconcini_2023}. 

\textbf{NGC~5643}: A Seyfert 2 spiral galaxy located at 17.3 Mpc. The double-sided ionization cone was examined by  \cite{Cresci_2015} who found that it was an asymmetric, blueshifted wing of the O\,{\sc  iii} emission line up to a projected velocity of $\sim$ $-450$ km $s^{-1}$, parallel to the low luminosity radio and X- ray jet, potentially collimated by a dusty structure around the nucleus. Additionally, in two star-forming clumps along the bar dust lane boundary \citep{Silk_2013}, \cite{Cresci_2015} revealed evidence of positive feedback induced by outflowing gas. These clumps were also characterized by modest CO~(2-1) emission \citep{Alonso2018}.

In this work we build upon the detailed study by \citet{Mingozzi_2019}, who disentangled the emission associated with galaxy disks from that of ionized outflows based on a kinematic decomposition of the emission-line profiles. 

They reduced the data by using the standard ESO pipeline, yielding datacubes with a spatial sampling of $0.2"$ and a spectral resolution ranging from R $\approx$ 1750 to 3750 across the optical range. The analysis was performed with custom Python scripts. The stellar continuum was modeled via pPXF \citep{pPXF_2004}, using single stellar population templates from \cite{Vazdekis_2010} and applied to Voronoi-binned spaxels to ensure an average continuum S/N of 50 for $\lambda$ < 5530 $\AA$. In Seyfert 1 nuclei (NGC~1365 and NGC~2992), additional components were included to account for broad-line emission and AGN continuum. After subtracting the fitted continuum on a spaxel-by-spaxel basis, the emission lines ([O\,{\sc iii}]$\lambda$$\lambda$4959,5007, [N\,{\sc ii}]$\lambda$6584, H$\alpha$, H$\beta$, [S\,{\sc ii}]$\lambda$6716, and [S\,{\sc ii}]$\lambda$6716, [O\,{\sc i}]$\lambda$$\lambda$6300,64) were fitted using MPFIT \citep{MPFIT_2009} with one to four Gaussian components, depending on the complexity of the line profile. For NGC~1365, we use binned velocity fluxes due to its low S/N ratio, whereas we employ smoothed velocity bin fluxes for the other galaxies.
Kinematics (velocity and dispersion) were tied across all lines in each component, while fluxes were left free, with fixed theoretical ratios for known doublets \citep[for more details see][]{Mingozzi_2019}. A reduced $\chi$$^{2}$ criterion was used to select the optimal number of components per spaxel. This approach allowed for a robust decomposition of the emission-line profiles, particularly in regions affected by non-Gaussian shapes such as galaxy centers and ionized outflows.

In Section 3 of \cite{Mingozzi_2019} analysis, they present  a method to disentangle the physical properties of the ionized gas associated with galaxy disks from those of the outflowing gas. By exploiting the spatial and spectral resolution of MUSE data, they perform a kinematic decomposition of the emission-line profiles into velocity bins \citep[see also][]{Venturi_2021}. The velocity channels near the stellar velocity trace the rotating disk component, while the high-velocity wings, typically at $\nu$ $>$ 150–250 km s$^{-1}$, are attributed to the outflow (their Fig. 1; Table \ref{table2}). This approach reveals that disk emission is generally associated with ring-like, spiral, or bar structures, whereas outflows usually appear as extended (bi)conical features on kiloparsec scales (their Fig. 2). The two kinematic components thus generally show distinct spatial distributions and different ISM properties (i.e., density, attenuation, ionization parameter, source of ionization), supporting a physical separation between rotating and outflowing gas. A comprehensive overview of the disparities between the disc and outflow gases for each of the nine galaxies under scrutiny is presented in Table \ref{table2}, expressing crucial insights into their unique characteristics and dynamics.

\section{AGN and star-forming region classification}
\label{sec:obs_classification}
\begin{figure*}
    \includegraphics[width=\linewidth]{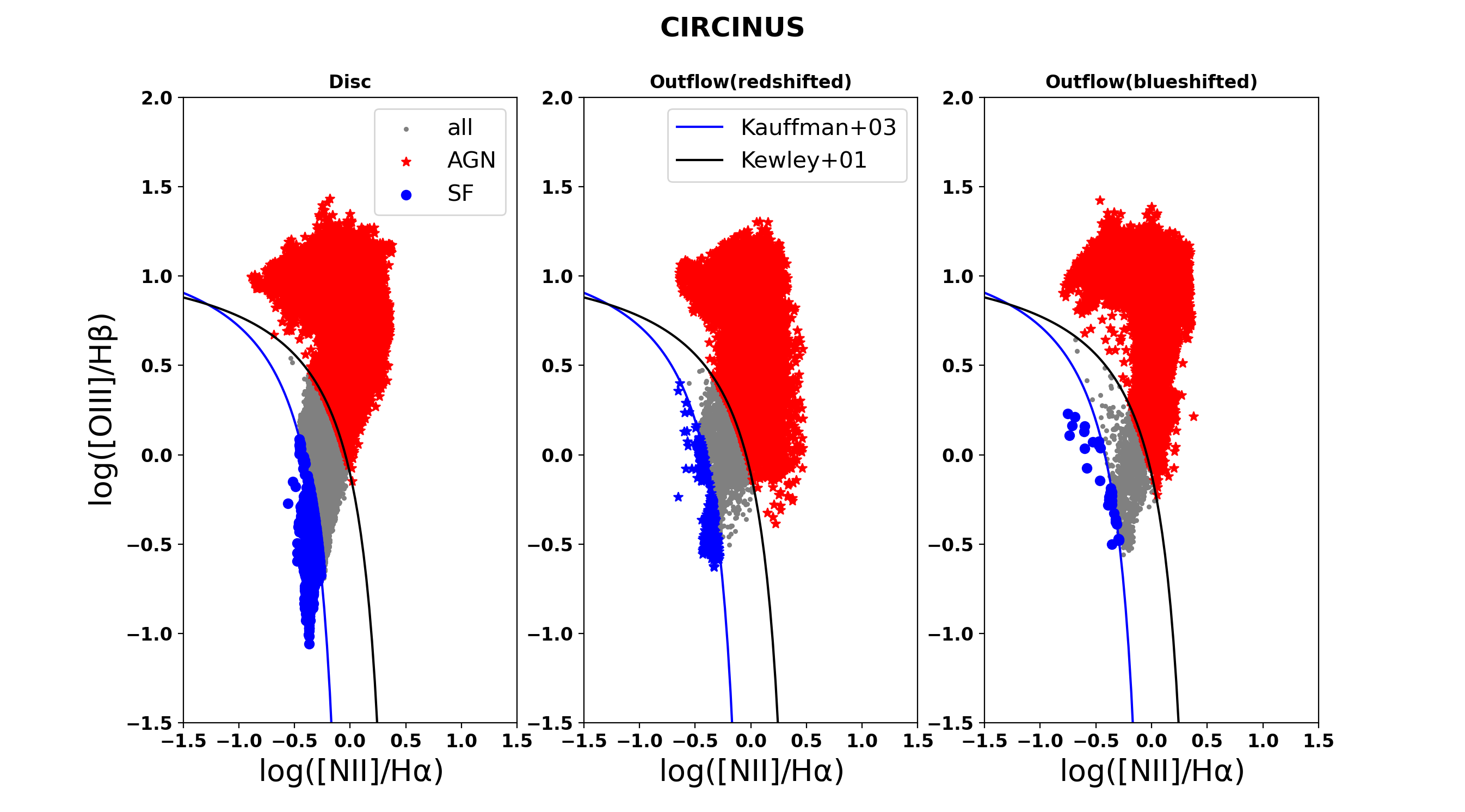}
    \includegraphics[width=\linewidth]{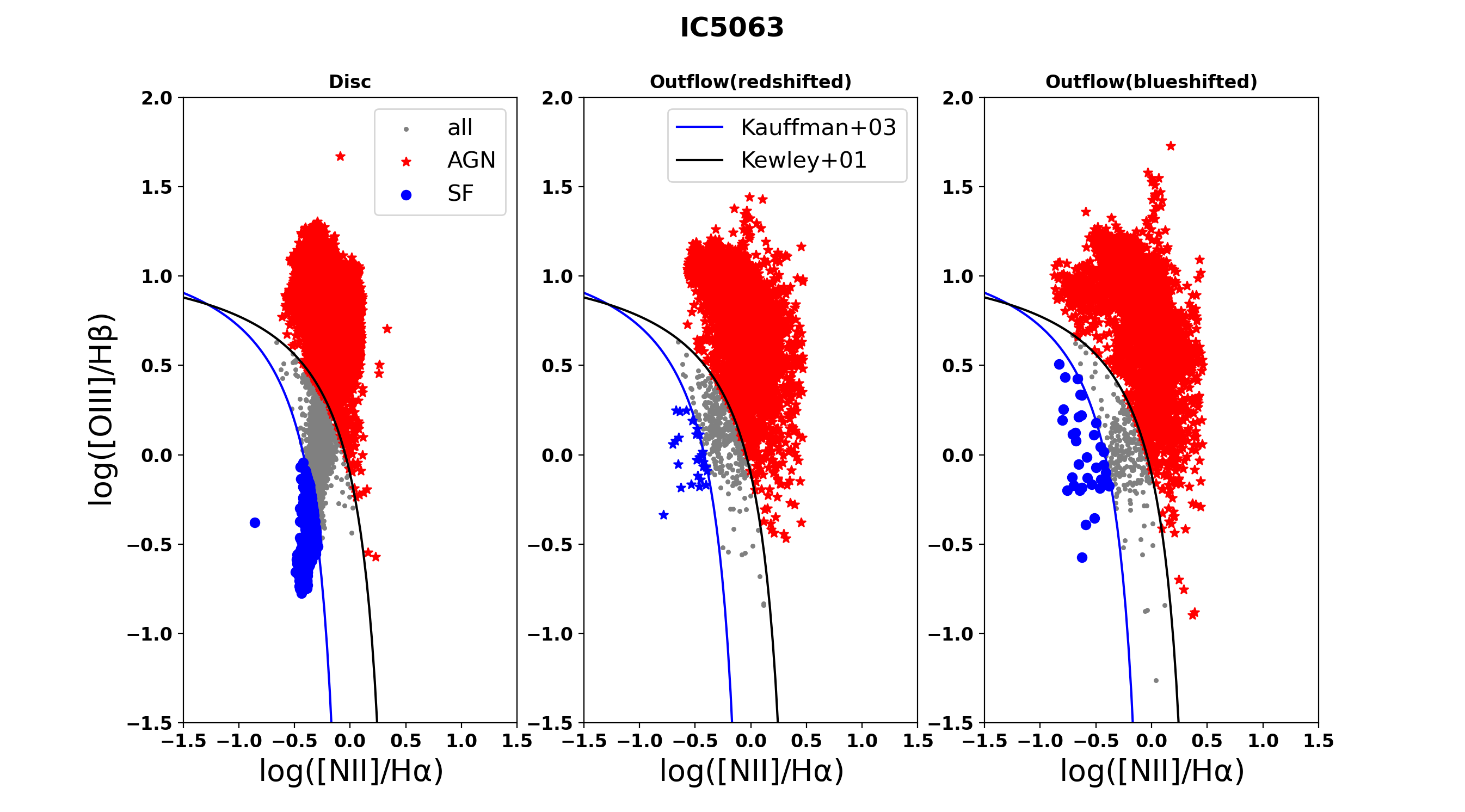}

    \caption{BPT diagnostic diagram for Circinus and IC~5063 to distinguish between AGN and SF ionised regions. The solid black line and the solid blue line show the discrimination between the two populations according to the boundaries from \citet{Kewley_bpt} and \citet{kauffmann_bpt}, respectively. Blue stars and red circles mark AGN- and SF-dominated regions, respectively, while small gray dots demonstrate the entire set of spaxel data.}
              
    \label{fig_bpt}
\end{figure*}

To classify regions in the nine MAGNUM galaxies as AGN-ionised or H\,{\sc ii}-ionised, we use the standard Baldwin–Phillips–Terlevich (BPT) diagrams \citep{bpt_main} applied to the analyzed spectral features of both disc and outflow components. The distribution of the bins in the [O\,{\sc iii}] 5007/H$\beta$ versus [N\,{\sc ii}]6584/H$\alpha$ diagram, also with the boundary lines between SF and AGN regions defined by \citet{Kewley_bpt} and \citet{kauffmann_bpt}, is shown in Fig.~\ref{fig_bpt}. The black demarcation line \citep{Kewley_bpt} is a theoretical upper limit on the location of SF galaxies in this diagram, obtained using a combination of photoionisation and stellar population synthesis models. It yields a conservative selection of AGN-dominated regions. \citet{kauffmann_bpt} revised this boundary line on the basis of observational evidence that SF regions and AGNs are distributed along two separate sequences. It yields a conservative SF region selection, which is shown in blue.

To avoid ambiguous classifications, we adopt the more conservative selection for both SF regions and AGNs, excluding from further consideration the bins located between the two lines. They could include regions with a mixture of ionising sources \citep[e.g.][]{Allen,rosa_interact_z} and are expected to have a mixed AGN-stellar emission as their ionizing source \citep[e.g.][]{Davis_2014} (see Figure~\ref{fig_bpt} for two examples; this is  fully demonstrated in appendix D of \cite{Mingozzi_2019}). Additionally, we exclude SF regions from the outflow, as our $Z_{\rm gas}$ assessments indicate larger metallicities than those associated with the AGN components of the outflow. They might be parts of the disc that have been inaccurately classified inside the outflow due to inadequate disc-outflow separation. Moreover, this might imply that the areas are not exclusively ionised by SF as there is an AGN contribution to the line emission, making the metallicity calibrators for SF regions unsuitable in these particular components.

\section{Estimating the gas-phase metallicity}
\label{sec:results_II}

To estimate $Z_{\rm gas}$ in each galaxy components, we use calibrations based on strong emission lines, commonly referred to as the \textit{strong-line method}. 
Computing element abundances using the direct method ($T_{e}$) is not possible as the MUSE wavelength coverage does not include the main temperature-sensitive emission lines [O\,{\sc  iii}]$\lambda$4363 and the doublet [O\,{\sc ii}]$\lambda$$\lambda$3727,3730. [N\,{\sc ii}]$\lambda$5755 and [S\,{\sc iii}]$\lambda$6312 are undetectable due to their exceedingly low S/N ratio in the MAGNUM galaxies. 
Also, we cannot calculate $Z_{\rm gas}$ from infrared nebular lines \citep{Z_gas_IR_2022} without the observation of a hydrogen recombination line within the infrared range.

Recently, \cite{HOMERUN_2024} established a new multi-cloud photoionisation approach (HOMERUN) to estimate $Z_{\rm gas}$ based on observed emission lines. This is a distinctive approach that the current version of their code does not yet include AGN photoionisation models. We are aware of the systematic variations in $Z_{\rm gas}$ that may occur when using strong-line approaches: differences of up to 0.6 dex for H\,{\sc ii} regions \citep[e.g.][]{ Kewley_2008_discrepancy,Sanch_discrepancy,Guerrero_2012_discrepancy} and up to 0.8 dex for AGNs can be found when comparing metallicities obtained using the strong-line method based on calibrations from different authors, particularly in the low-metallicity regime (less than 8.5, e.g.,\citealt[][]{Dors_2020_b_discrep}). As discussed in detail in \cite{Armah_2023}, the different calibrations employ distinct emission line ratios which are sensitive to varying physical conditions such as ionization parameter, density of gas, and metallicity, leading to divergent gradient estimates.

To estimate $Z_{\rm gas}$ based on the \textit{strong-line method}, we utilize the emission line intensities of [O\,{\sc iii}]$\lambda$5007, [N\,{\sc ii}]$\lambda$6584, H$\alpha$, H$\beta$, [S\,{\sc ii}]$\lambda$6716, and [S\,{\sc ii}]$\lambda$6716. We consider S/N$>$5 for each emission line. Bins with no detection of one or more of these emission lines were discarded. To estimate $Z_{\rm gas}$ for regions dominated by AGN and SF, we consider two distinct calibration relations.
For SF regions, we examine the O3N2 methodology by \citep{Alloin_first_o3n2} which is developed by \citep{pettini}:
\begin{equation} \label{pettini}
\begin{split}
Z_{\rm gas} = 8.073 - 0.32\times({\rm O3N2})
\end{split}
\end{equation}
Where O3N2 is defined as:
\begin{eqnarray} \label{pettini}
{\rm O3N2} & = & {([{\rm O}\,\textsc{iii}]\lambda5007/{\rm H}\beta)/ ([{\rm N}\,\textsc{ii}]\lambda6584/{\rm H}\alpha}).
\end{eqnarray}

We emphasize that the observations are not included in the data release. As a result, individual spaxels likely contain more than one H\,{\sc ii} region, and the metallicities presented here should be regarded as spatially averaged values rather than measurements of single H\,{\sc ii} regions.

To calculate $Z_{\rm gas}$ in each  spaxel containing  AGN regions, we use the relation from \citet{Storchi-Bergmann_1998}. We first calculate $Z_{\rm gas}$ from the optical emission-line ratios of AGN that are valid for $Z_{\rm gas}$ in the range $8.4\leq Z_{\rm gas}\leq 9.4$. The values of $Z_{\rm gas}$ computed from these calibrations vary by $\sim0.1\,{\rm dex}$. The metallicity value is corrected in order to take electron density ($n_{\rm e}$) effects into account via:
\begin{equation} \label{berg1}
Z_{\rm gas} = {Z_{\rm int} - (0.1\times \log_{10}(n_{e} /300\,{\rm cm}^{-3}))}
\end{equation}
in which\begin{equation} 
\label{berg2}
\begin{split}
Z_{\rm int} = 8.34 +(0.212\times {\rm N2})
-(0.012 \times {\rm N2}^{2})-(0.002\times {\rm R3})
\\+ (0.007\times ({\rm N2}\times R3))- (0.002\times {\rm N2}^{2}\times {\rm R3})\\ + (6.52\times (10^{-4}\times {\rm R3}^{2}))
+ (2.27\times 10^{-4}\times ({\rm N2}\times {\rm R3}^{2})) \\+ (8.87\times 10^{-5} \times ({\rm N2}^{2} \times {\rm R3}^{2}))
\end{split}
\end{equation}
and where N2, S2, and R3 are defined as:

\begin{eqnarray}
{\rm N2} & = &  {[{\rm N}\,\textsc{ii}]\lambda6584/{\rm H}\beta},\\
\nonumber\\
\label{mirko_4}
{\rm S2} &=& ({[{\rm S}\textsc{ii}]\lambda6716+[{\rm S}\textsc{ii}]\lambda6730)/{\rm H}\beta},\\
\nonumber\\
\label{mirko_5}
{\rm R3} &=& {[{\rm O}\textsc{iii}]\lambda5007/{\rm H}\beta}.
\end{eqnarray}

\begin{figure*}[t]
  \centering
  \begin{tabular}{cc}
    \includegraphics[width=0.51\textwidth]{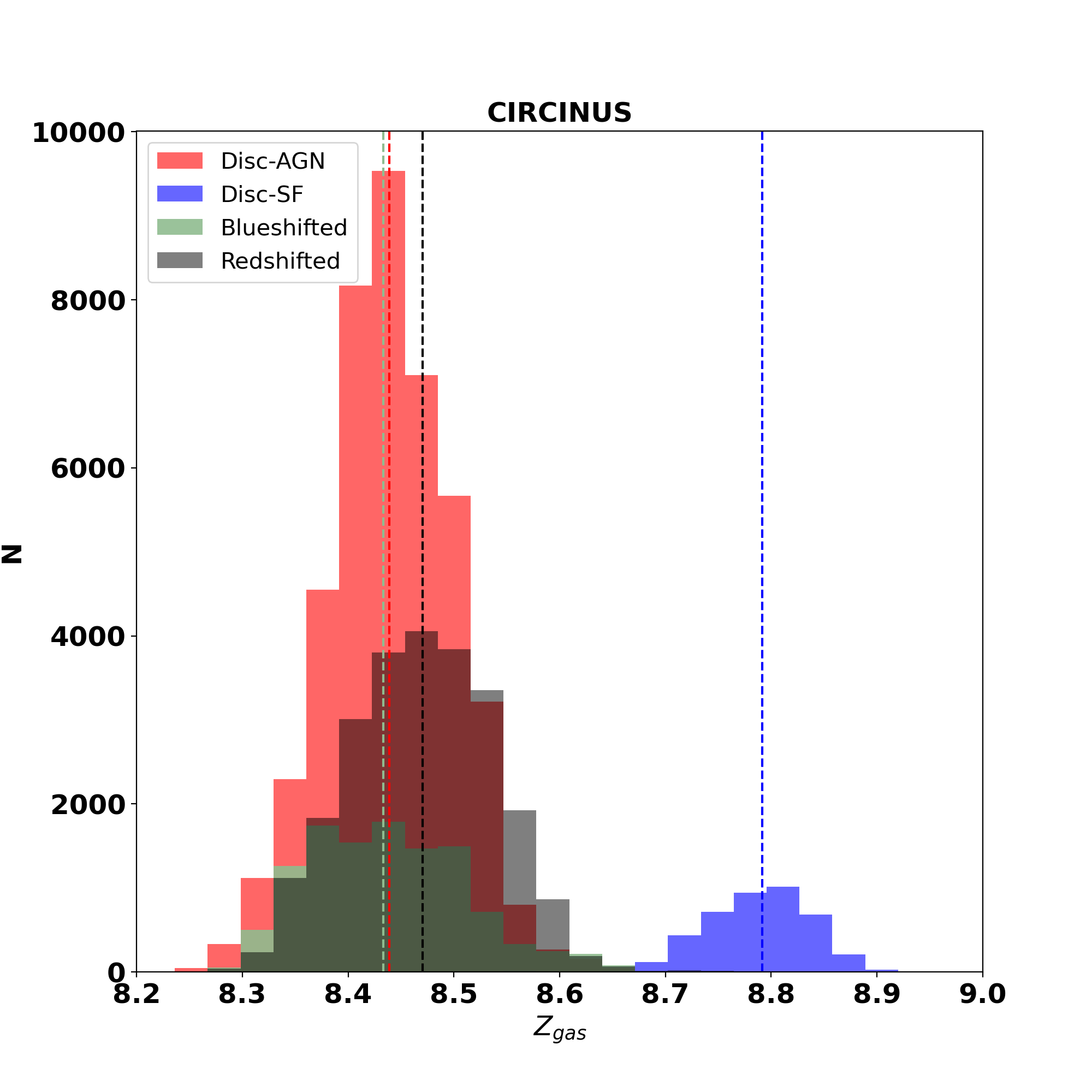}
  \includegraphics[width=0.51\textwidth]{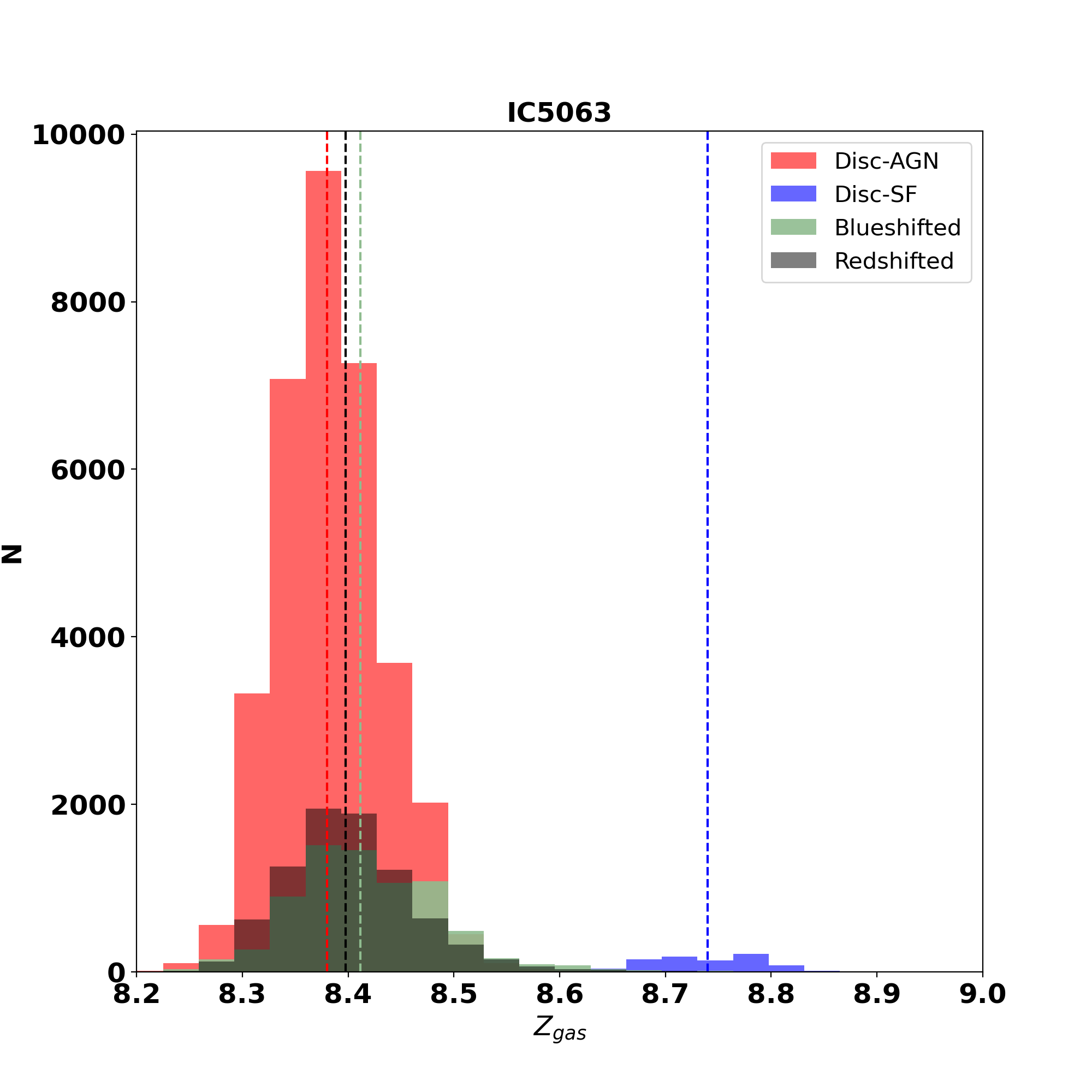}\\
  \end{tabular}
  \caption{Distribution of $Z_{\rm gas}$  for the AGN parts in disc(red), redshifted (black) and blueshifted (green) outflow. SF-regions in disk is plotted in blue color. The vertical dashed lines show the median value of $Z_{\rm gas}$  for each of the components in the same color. The $Z_{\rm gas}$} for CIRCINUS, and IC~5063 are shown here while other galaxies are demonstrated in \ref{sec:appB}. 
  \label{fig_histograms_zgas}
\end{figure*}

To estimate $n_{\rm e}$ we assume a temperature of $T_{e}$ = $10^4$K and measure the optical  [S\,{\sc ii}]$\lambda$6716 /[S\,{\sc ii}]$\lambda$6731 ratio, converting it to an electron density using the  \citet{Osterbrock2006} model via Pyneb \citep{Pyneb}. The results are shown in the  Fig~\ref{fig_histograms_zgas} and Appendix A.

It is imperative to stress that this calibration relies on one--dimensional photoionization models. In the calibration of \citet{Storchi-Bergmann_1998}, and similar approaches \citep[e.g.][]{Zhu_2024}, the ionizing source is centralized, and computations are concluded over the nebular radius, anticipating emission line fluxes. This may provide higher uncertainties when deriving $Z_{\rm gas}$ from spatially resolved observational data, but not from integrated data.

\section{Results}
\label{radial_Z}

\begin{table*}
\caption{General information on MAGNUM galaxies}
\begin{tabular}{ccccccccc}

    \hline
    %\multirow{2}{*}{Name} \\
     Name & AGN type & $D_{l}$ [Mpc]$^{1}$ & BH mass[$M_{\sun}$]$^2$ & morphology/type & interaction$^3$ & $1"$ [pc]\\% &Torus\\
    \hline
    Centaurus A& Seyfert 2 & 3.82 & $6.5\times10^7a$&Lenticular/early type& yes $j$& 18.5\\%& yes $p$\\
    Circinus & Seyfert 2 & 4.2 & $\approx10^ {6.23}b$ & spiral& no & 20.4 \\%&yes $q$\\
    IC~5063  & Seyfert 2 & 45.3 & $5.5\times10^7c $ & early type&yes $l$& 219.6 \\%&yes $r$\\
    NGC~1068 & Seyfert 2 & 12.5 & $\approx10^{7.23}d$ & spiral& yes $m$& 69 \\%&yes $s$\\
    NGC~1365 & Seyfert 1.8 & 18.6 & $2\times10^6e$ & spiral& no& 90.2 \\%&yes $t$\\
    NGC~1386 & Seyfert 2 & 15.6 & $\approx10^{7.24}f$ & Sb/c spiral&$-$& 76\\% &yes $u$\\
    NGC~2992 & Seyfert 1.9 & 31.5 & $\approx10^{7.72}g$ & spiral& yes $o$& 150\\% &yes $v$\\
    NGC~4945 & Seyfert 2 & 3.7 & $\approx10^{6.5}h$ & spiral& no& 18 \\%&yes $w$\\
    NGC~5643 & Seyfert 2 & 17.3 & $\approx10^{6.4}i$ & spiral & no& 84\\% &yes $x$\\
    \hline

\end{tabular}
\label{table1}
\tablefoot{1) Luminosity distance 2)Black Hole(BH) mass 3)To determine if a galaxy reveals interactions with its surroundings. a)\citet{bh_centaurus} b)\citet{bh_circinus} c,d,e,f,g)\citet{bh_ic5063_1068_1365_1386_2992} h)\citet{bh_ngc4945} i)\citet{bh_NGC5643} j)\citet{merging_centaursa} l)\citet{merging_ic5063}  m)\citet{merging_1068} o)\citet{merging_2992} %p)\citet{torus_centaurusa} %q)\citet{torus_circinus} %r)\cite{torus_1_ic5063,torus_2_ic5063} 
%s)\citet{torus_ngc1068} %t)\citet{torus_ngc1365} %u)\citet{torus_ngc1386} %v)\citet{torus_NGC2992} %w)\citet{torus_ngc4945} %x)\citet{torus_5643}
}

\end{table*}

\begin{table}

\caption{MAGNUM galaxies components (all velocities are in $[km s^{-1}]$)}.

\hskip-0.6cm
\begin{tabular}{cccc}

    \hline
    %\multirow{2}{*}{Name} \\
     Name & Disc &outflow (Red) & outflow (Blue)\\
    \hline
    Centaurs A & -100 $<\nu<$+100&$\nu > +150$ & $\nu<-150$\\
    Circinus &-150 $<\nu<$+150& $\nu > +200$ & $\nu<-200$ \\
    IC~5063   &-150 $<\nu<$+150 & $\nu > +200$ & $\nu<-200$\\
    NGC~1068  & -150 $<\nu<$+150& $\nu > +250$ & $\nu<-250$\\
    NGC~1365  & -100 $<\nu<$+100& $\nu > +150$& $\nu<-150$\\
    NGC~1386  & -100 $<\nu<$+100& $\nu > +150$& $\nu<-150$\\
    NGC~2992  & -150 $<\nu<$+150& $\nu > +200$ & $\nu<-200$\\
    NGC~4945  & -150 $<\nu<$+150& $\nu > +200$ & $\nu<-200$\\
    NGC~5643 & -150 $<\nu<$+150& $\nu > +200$ & $\nu<-200$\\
    \hline

\end{tabular}
\label{table2}
\tablefoot{Associated velocity channels around the core of the lines in the fitted line profile with the disc component and the total of the outflow redshifted and blueshifted velocity channels for each galaxy from \cite{Mingozzi_2019}.}
\end{table}

\subsection{Radial variations of $Z_{\rm gas}$}

\begin{figure*}
\vspace{-0.1cm}    %\includegraphics[width=\linewidth]{figures/last_plot_1.pdf}
\includegraphics[width=\linewidth]{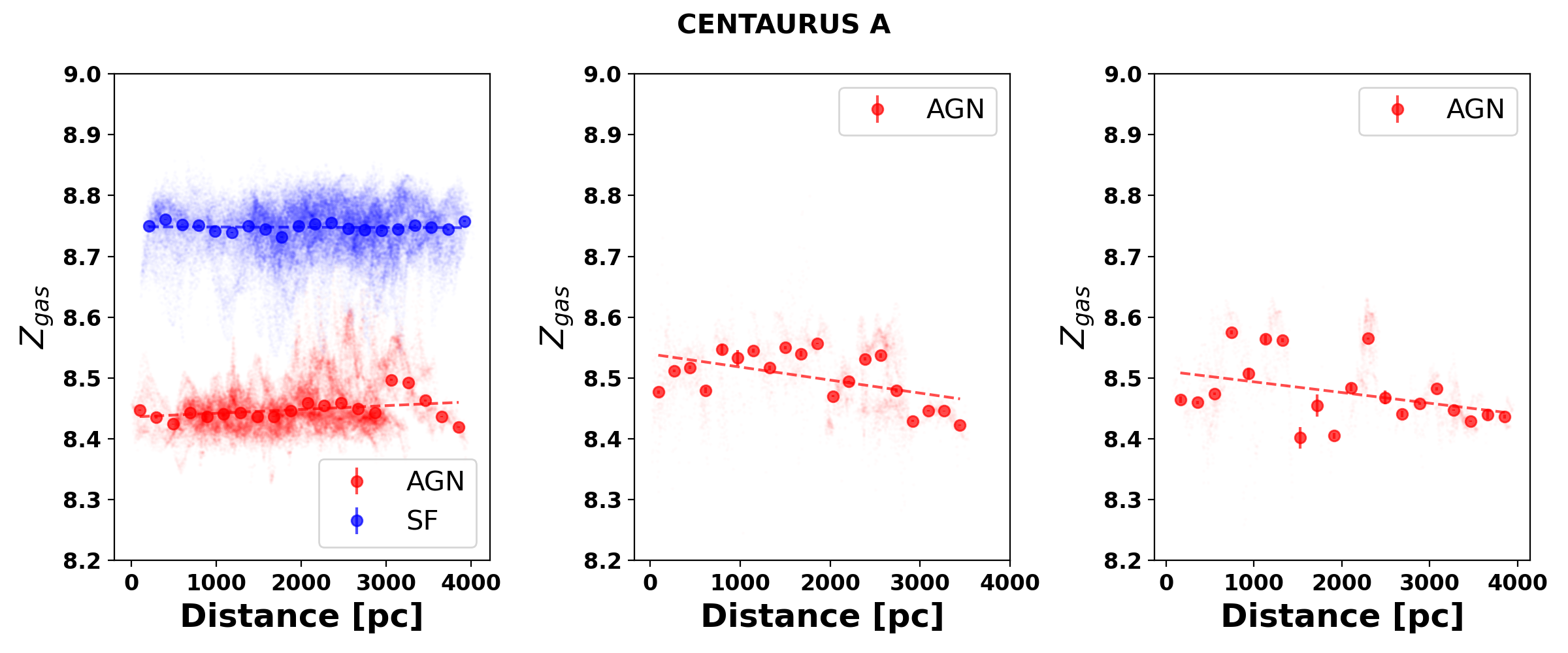}
\includegraphics[width=\linewidth]{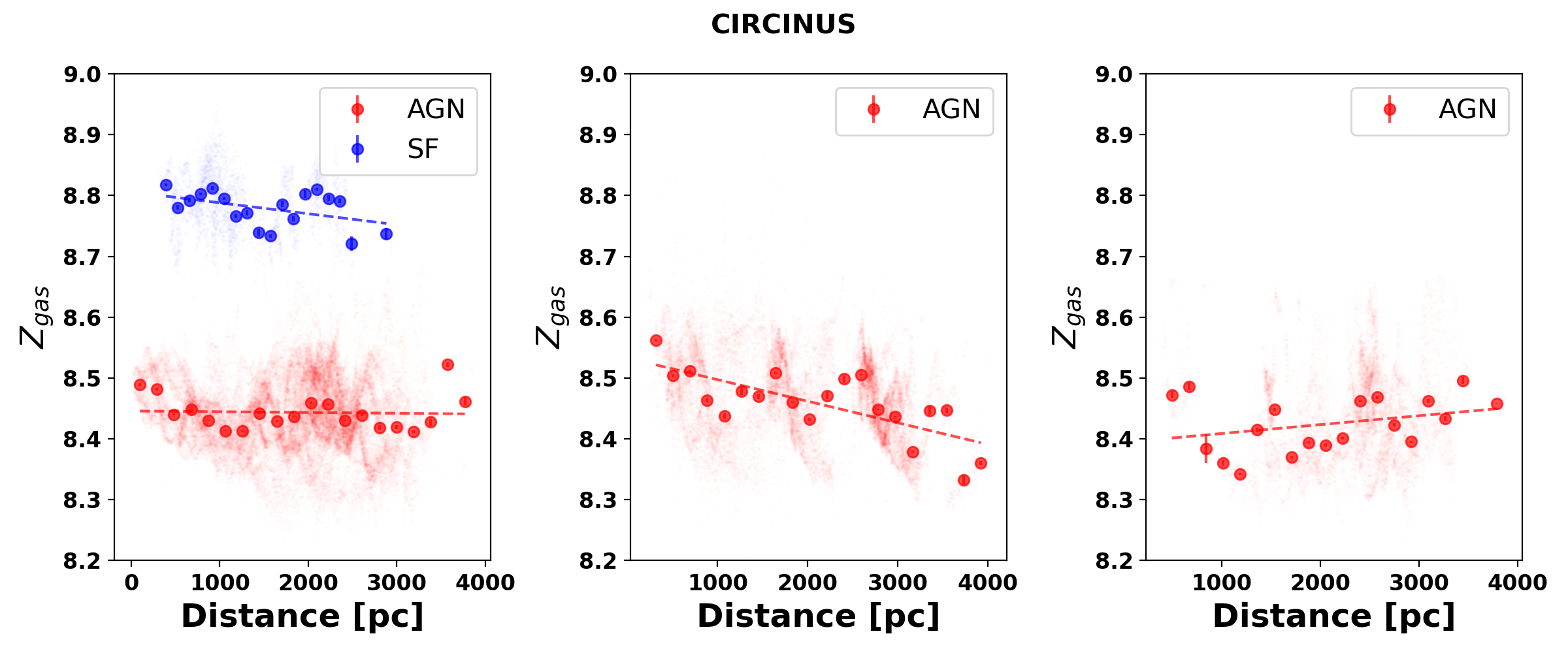}
\includegraphics[width=\linewidth]{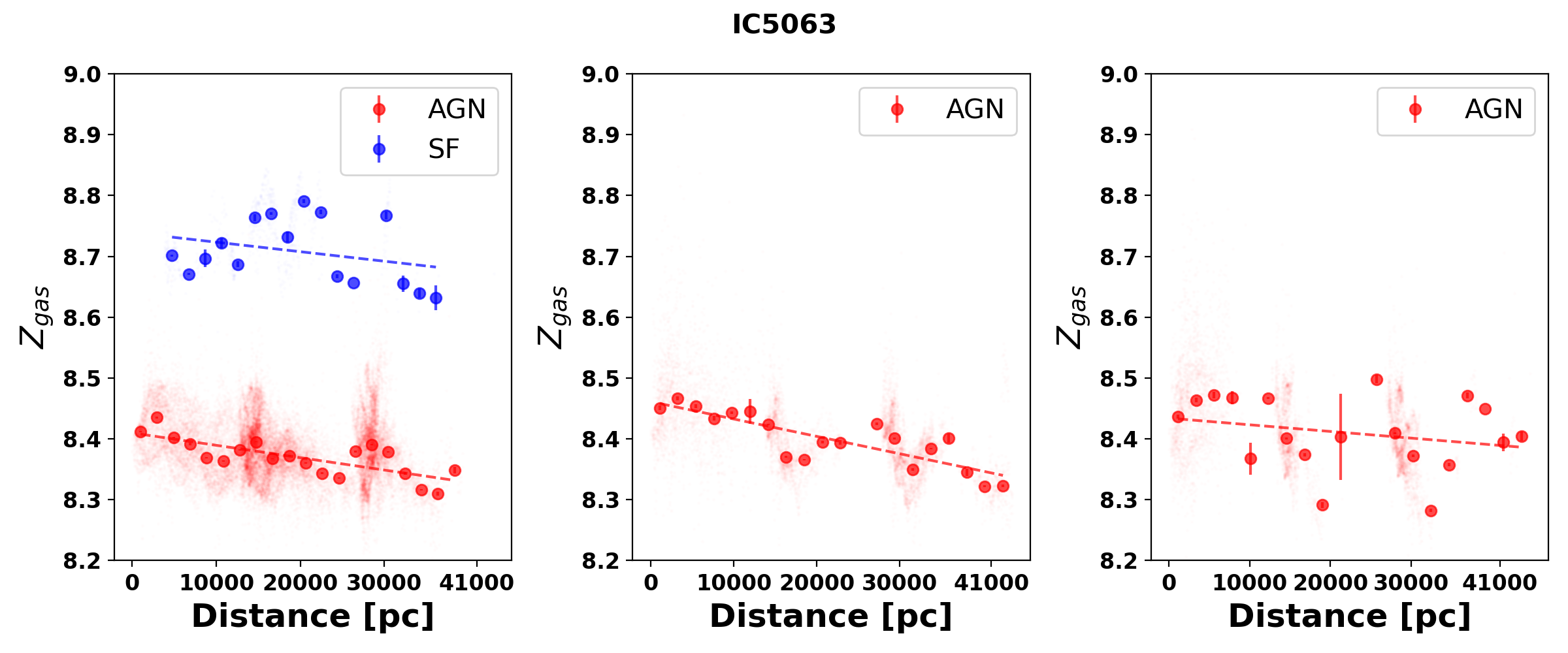}

    \caption{
    Centaurs A, Circinus, and IC~5063 metallicity gradient profiles for disc (left column), redshifted (middle column) and blueshifted (right column) outflow of MAGNUM galaxies using the \citet{Storchi-Bergmann_1998} and \citet{pettini} metallicity relations for AGNs (red circles) and SF regions (blue circles), respectively, with the radius (in pc) given on the bottom axis. Overplotted are binned linear fits to the radial metallicity gradient in AGNs (dotted red lines) and SF regions (dotted blue lines). The bigger dots mark the binned median values, and error bars represent the scatter in the data within each bin.}
              
    \label{fig_zr_3_1}
\end{figure*}

\begin{figure*}
    \includegraphics[width=\linewidth]{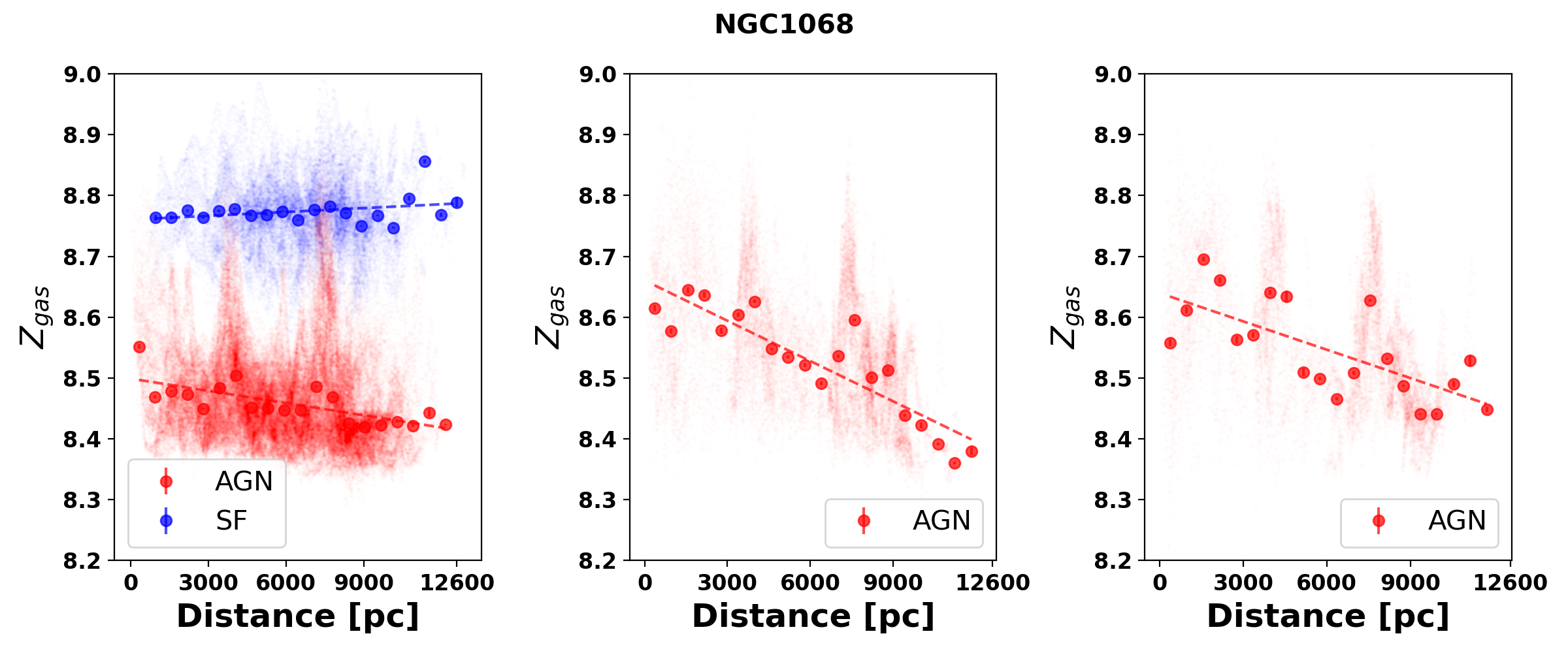}

    \includegraphics[width=\linewidth]{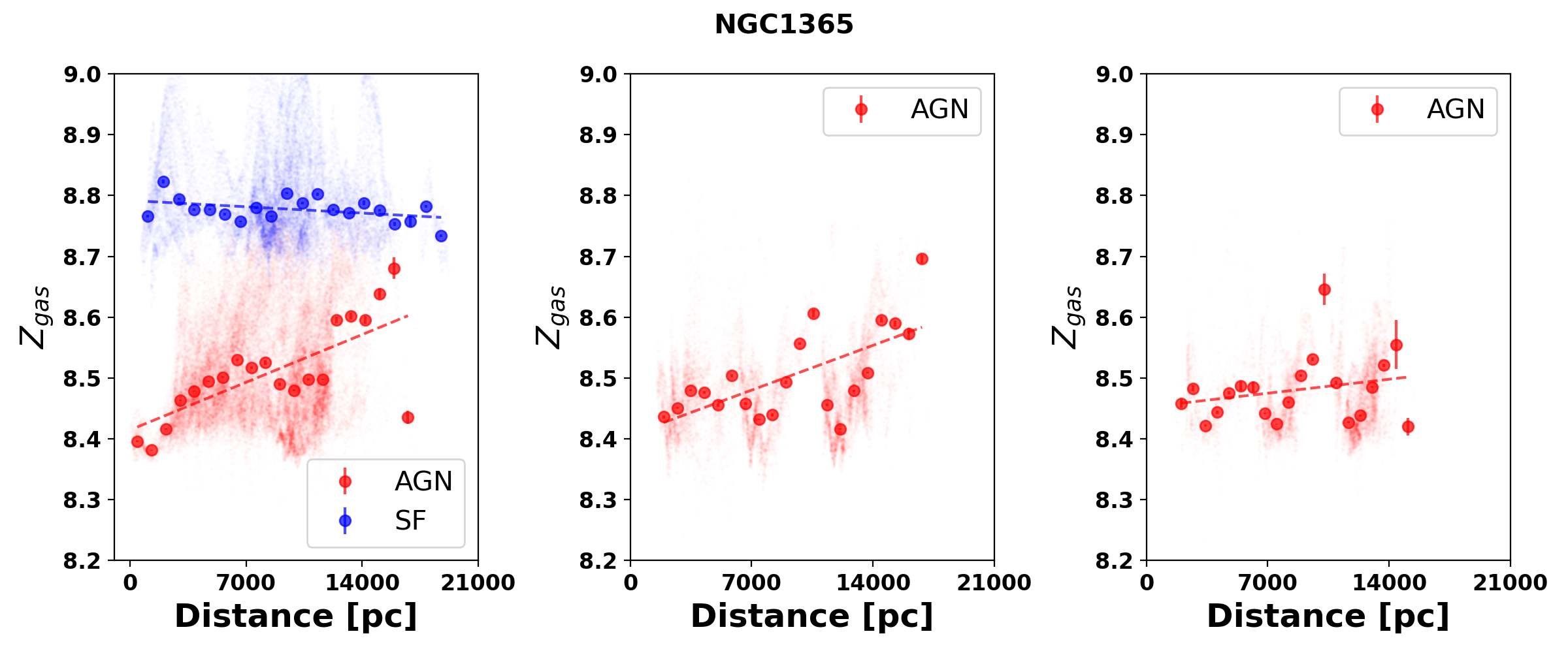}
    \includegraphics[width=\linewidth]{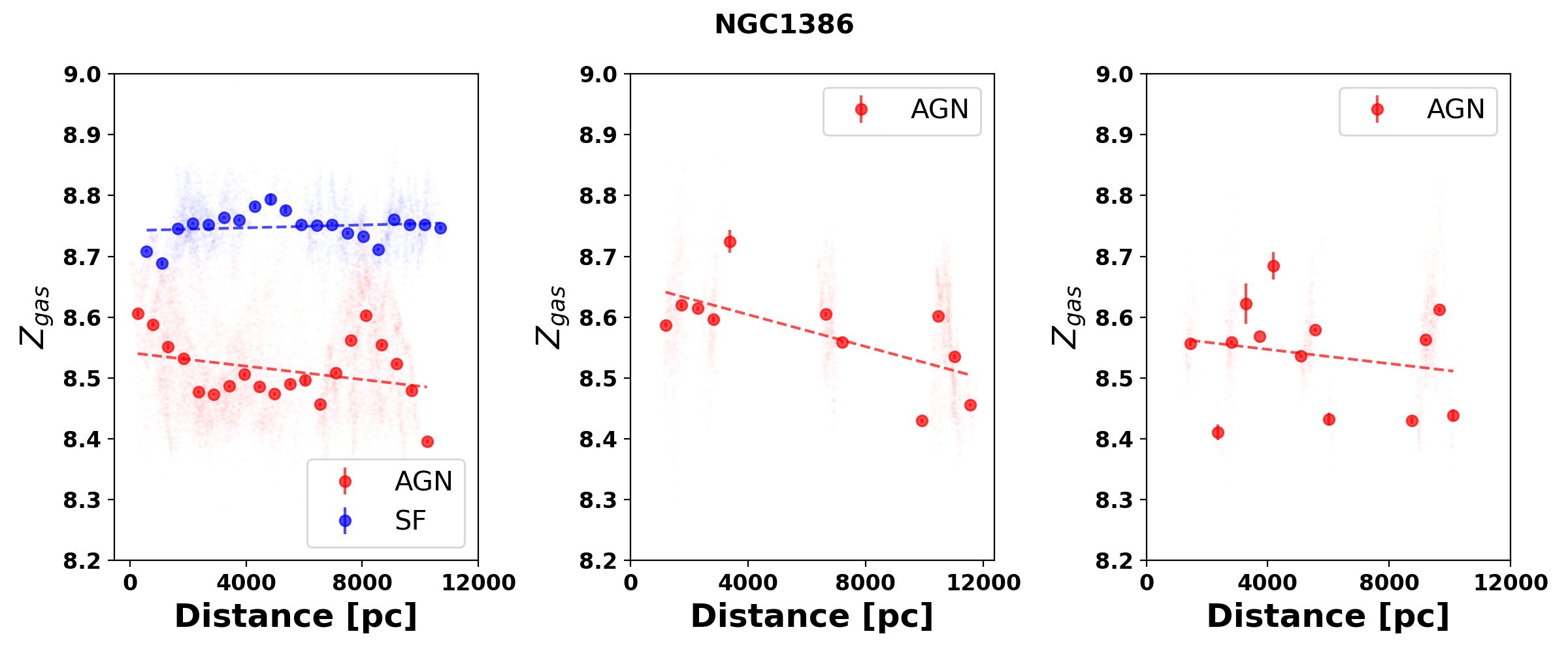}
    
    \caption{Same as Fig \ref{fig_zr_3_1}, but for NGC~1068, NGC~1365, and NGC~1386.}
              
    \label{fig_zr_3_2}
\end{figure*}

\begin{figure*}
    \includegraphics[width=\linewidth]{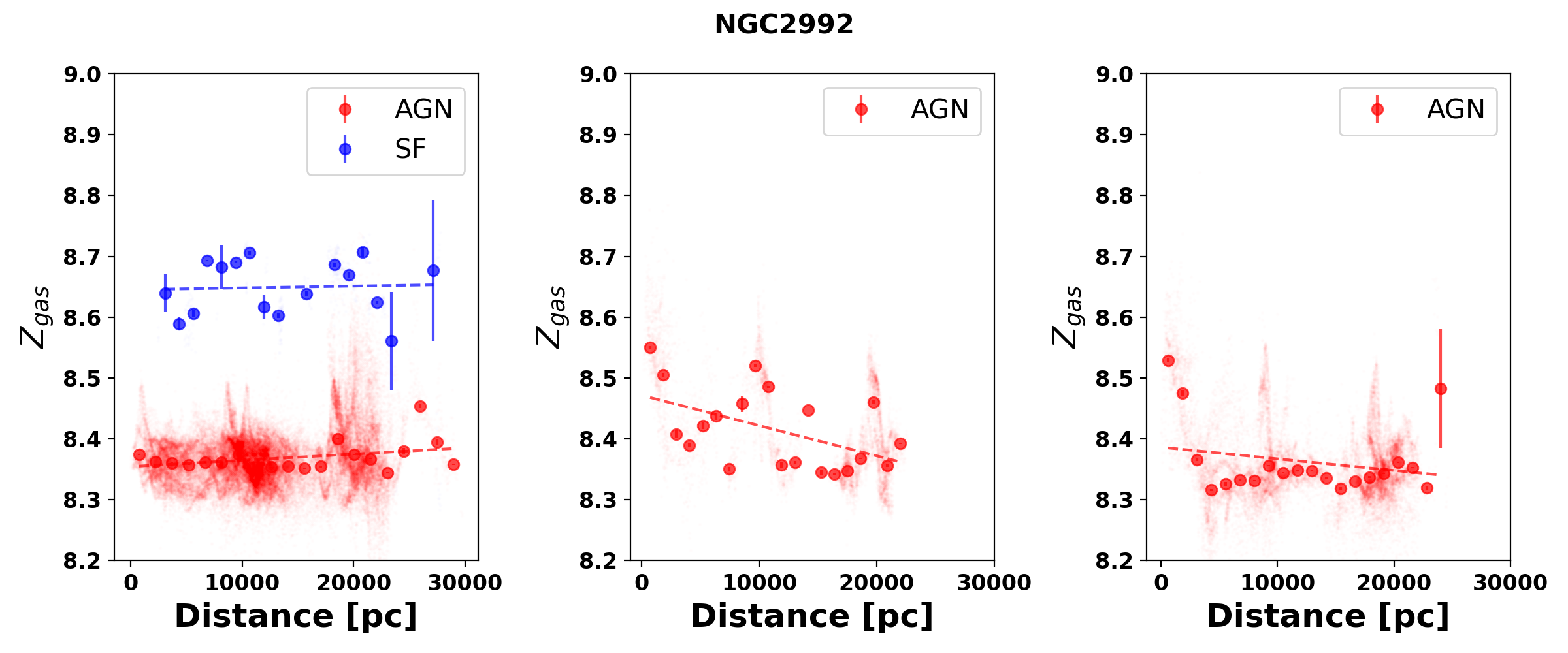}
    \includegraphics[width=\linewidth]{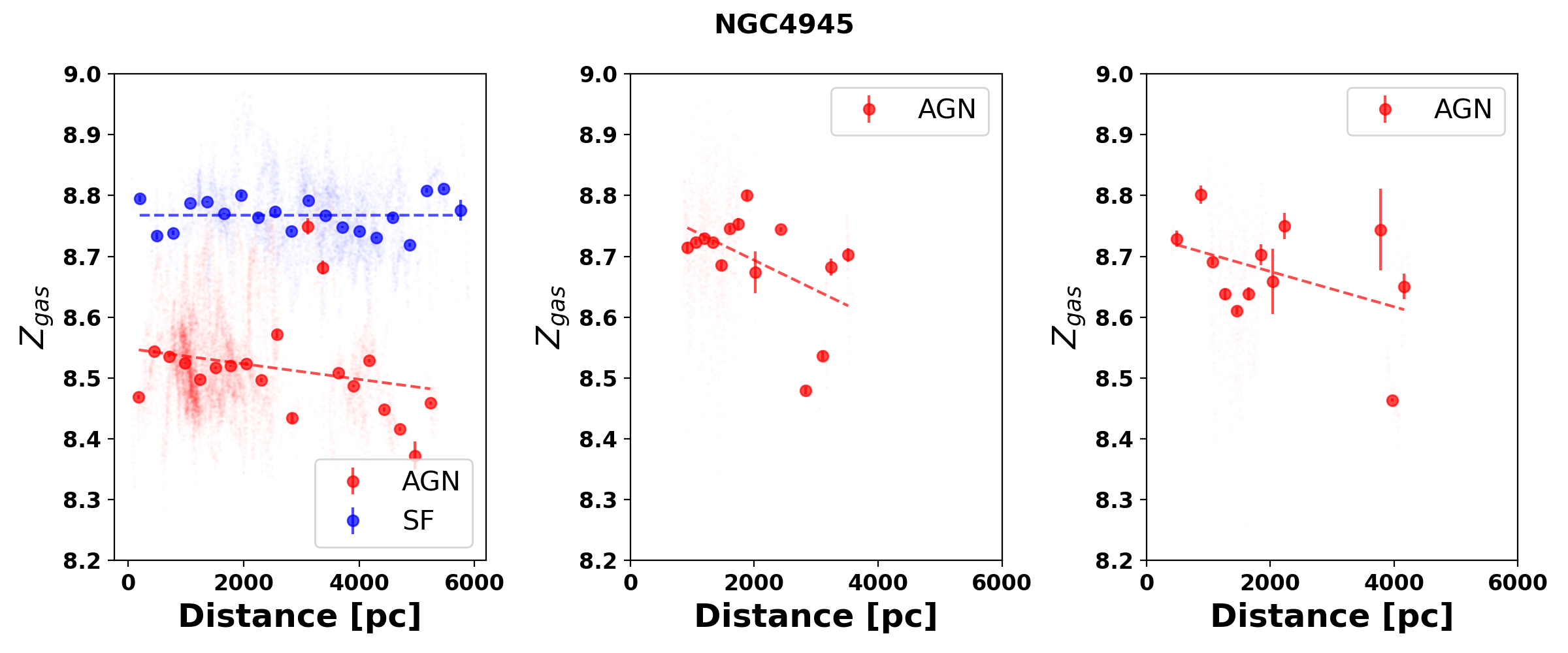}
    \includegraphics[width=\linewidth]{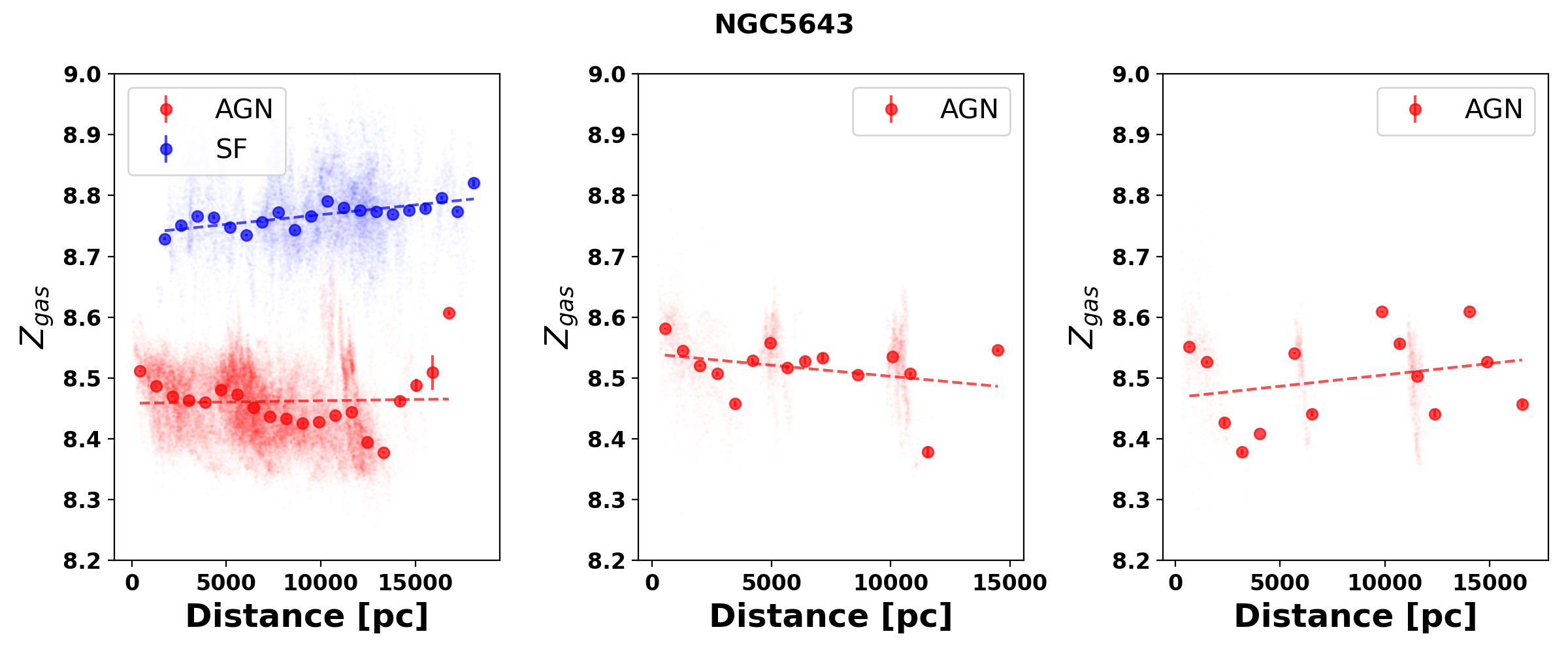}
    \caption{Same as Fig \ref{fig_zr_3_1}, but for NGC~2992, NGC~4945,NGC~5643.}
              
    \label{fig_zr_3}
\end{figure*}

We derive radial $Z_{\rm gas}$ distribution based on strong line calibration relations. To calculate variations in $Z_{\rm gas}$ relative to the galactic center, we estimate the projected distance from the [O\,{\sc iii}]$\lambda$5007 emission spatial peak. We define the [O\,{\sc iii}]$\lambda$5007 peak as the galaxy center by determining the pixel with the largest flux, subsequently calculating the projected distance of each pixel from this center using the Euclidean distance scaled by the pixel size. We derive radial oxygen abundance profiles for the disc, redshifted outflow, and blueshifted outflow components of MAGNUM galaxies, as shown in Figs.\ref{fig_zr_3_1},\ref{fig_zr_3_2}, and \ref{fig_zr_3}. We employ a binned linear regression approach to fit the radially averaged $Z_{\rm gas}$ as a function of radial distance. In Table \ref{table3}, we list the radial $Z_{\rm gas}$  behavior for each galaxy component. Our main redshifted and blueshifted outflows demonstrates no particular variation from the negative radial $Z_{\rm gas}$  gradients observed in non-active galaxies, suggesting that AGN feedback does not particularly lead the redistribution of metals on galactic scales. This aligns with earlier studies indicating that AGN-driven outflows tend to be collimated, anisotropic, and episodic, hence limiting their ability to homogenize the chemical composition of the ISM \citep[e.g.][]{Harrison2018, Venturi_2021}.The observed gradients are more effectively explained by an inside-out evolution model, wherein earlier and sustained star formation in the center regions results in elevated metallicities compared to the outskirts \citep[e.g.][]{Perez2013,Sanchez2014}. The results indicate that, for mainly redshifted and blueshifted outflows, the impact of AGN feedback on radial $Z_{\rm gas}$ may not be the main strong reason to redistribute the metal contents. We found that the negative trend in the radial $Z_{\rm gas}$ in both redshifted and blueshifted outflows in the MAGNUM survey is incapable of redistributing centrally enriched material to the outer regions of their host galaxies. The kinetic power of these outflow components is inadequate to displace the metals from the central area. This is reflected in the observed steep negative radial metallicity gradients, with no compelling evidence for chemical enrichment beyond the inner regions. Similar conclusions have been drawn in recent integral field spectroscopic studies, which demonstrate that while outflows can perturb the circumnuclear interstellar medium, they typically do not possess the energy required to drive galaxy-scale transport of metals \citep[e.g.][]{Venturi_2018,Mingozzi_2019}.
This finding is also supported by studies of more extreme systems: in the ultraluminous infrared galaxies (ULIRGs), outflows show greater mechanical output and can contribute to chemical enrichment beyond the disk \citep{Rupke2005}. Simulations further indicate that only high-luminosity AGN can launch outflows capable of varying metallicity profiles on large scales \citep[e.g.][]{Nelson2019}. Therefore, in these low-luminosity AGN, the feedback appears spatially limited and chemically ineffective at the scale of the entire galaxy.

Additionally, blueshifted outflow regions of Circinus and NGC~5643 also have a positive radial $Z_{\rm gas}$. Several nearby AGNs have been shown to exhibit positive gradients, such as NGC~7130 \citep{Amiri_2024} and around 20 percent of the Seyfert galaxies from the SDSS-IV MaNGA survey \citep{Nascimento_manga}. 

NGC~1365 is the only galaxy in the MAGNUM survey whose redshifted and blueshifted outflow components, as well as its disc AGN component, exhibit a positive radial gradient in $Z_{\rm gas}$, as shown in Fig. \ref{fig_zr_3_2}.

Table \ref{table1} highlighted that NGC~1365, Circinus, and NGC~5643 show also no evidence of interactions with their nearby galaxies. In such a case, metal-enriched outflows may transport heavy elements from the central regions to the outskirts of the galaxy through the absence of external environmental influence. Given the substantial energy required to transport this gas to its observed position, AGN-driven physical properties are considered the most plausible mechanism for such metal content displacement \cite[e.g.][]{Simionescu_ff,Pope,Kirkpatrick_ff}. This aligns with the observational evidence supporting the presence of metal-rich outflows driven by AGN-powered winds and/or jets \citep{McNamara_2012,Amiri_2024}. 

Interpreting the ionization sources within galactic discs is more complicated because those regions are often ionized by both AGN and SF. The AGN radiation can ionize parts of the disc, especially if the AGN outflow axis is aligned close to the disc plane. However, an alternative is that some regions identified as AGN-ionized are actually outflow components that were misclassified as disc components due to limitations in kinematic analysis, particularly difficulties in accurately measuring the intensity of the [O III] line associated with the disc.
Regions of the disc ionized by an AGN typically show a positive radial gradient in $Z_{\rm gas}$. By contrast, SF and AGN regions within the disc exhibit a negative radial $Z_{\rm gas}$ gradient in 60 percent, while 40 percent of our AGN-ionized disc regions also show this negative trend. This negative gradient in the disc aligns with the idea that galaxies have relatively uniform gas accretion histories, where metal-poor inflows and outflows primarily influence the outer parts of galaxies. This, combined with the inside-out evolution of galaxies, naturally leads to a negative metallicity gradient \citep[e.g.][]{Boardman}. Conversely, in simpler scenarios where galaxies form and evolve in dense environments with significant gas flows \citep[e.g.][]{Somerville}, the radial $Z_{\rm gas}$ gradient can be inverted (positive). Consequently, the combination of these scenarios can provide a dual behavior (positive and negative) for radial $Z_{\rm gas}$ in the disk.

\begin{table}
    \centering
\begin{tabular}{ccccc}
        \hline
        \multirow{2}{*}{{Name}} & \multicolumn{2}{c}{{Disc}} & \multirow{2}{*}{{Outflow (Red)}} & \multirow{2}{*}{{Outflow (Blue)}} \\ 
        \cline{2-3}
        & {AGN} & {SF} &  &  \\ 
        \hline
        Centaurs A & + & - & - & - \\ 
  
        Circinus   & - & - & - & + \\ 

        IC~5063     & - & - & - & - \\ 

        NGC~1068    & - & + & - & - \\ 

        NGC~1365    & + & - & + & + \\ 

        NGC~1386    & - & + & - & - \\ 

        NGC~2992    & + & + & - & - \\ 

        NGC~4945    & + & - & - & - \\ 

        NGC~5643    & + & + & - & + \\ 
        \hline
        \hline

\end{tabular}
    \caption{The general behavior of $Z_{\rm gas}$ radial variation. Positive and negative behaviors are denoted by + and - signs, respectively. }
    \label{table3}
\end{table}

\section{Discussion and Conclusions}
\label{sec:conclusion}
We investigate the unresolved issue of the radial distribution of gas-phase metallicities in AGN by exploring the disc and outflow gas components discovered in the MAGNUM survey by \cite{Mingozzi_2019}. We differentiate between SF and AGN dominated regions by using the conventional BPT diagnostics. Depending on the type of emission lines that may be detected in both AGN and SF regions, many studies have proposed varying strong line ratios to characterize the metallicities \cite[e.g.][]{Carvalho_agn_calibration,Storchi-Bergmann_1998, perez_montero,curti_stack}. Depending on the location in the BPT diagram, we compute gas-phase metallicities based on either the \cite{Storchi-Bergmann_1998} and \cite{pettini} calibration relations. We subsequently analyze the radial distribution of gas-phase metallicity for disc and   redshifted/ blueshifted outflow.

In the outflow (redshifted and blueshifted) regions, we discover a predominantly negative radial trends, indicating a decrease in $Z_{\rm gas}$ through the outer regions. This indicates that the total radial $Z_{\rm gas}$ follows an inside-out scenario. Our analysis reveals that ionized outflows (both redshifted and blueshifted) in MAGNUM galaxies do not carry sufficient energy to drive metal-rich material from the nuclear regions to the outskirts of their host galaxies. The persistence of steep negative radial metallicity gradients suggest that such outflows are ineffective for  chemical redistribution. This supports a picture in which AGN-driven feedback in typical Seyfert galaxies is confined to the central regions and does not significantly impact the chemical evolution of the outskirts of a galaxy. These results align with both observational \citep[e.g.][]{Venturi_2018,Mingozzi_2019} and numerical simulations \citep[e.g.][]{Nelson2019} studies that contrasts with findings in more energetic environments such as ULIRGs \citep{Rupke2005}, where outflows can influence the circumgalactic medium. 

We also observe a dual behavior in the disc component of both AGN and SF dominated regions, showing both positive and negative metallicity gradients. The negative trend corresponds with the concept that galaxies possess generally consistent gas accretion histories, where metal-poor inflows and outflows predominantly affect their outer regions.  This, combined with the inside-out evolution of galaxies (where central regions form stars and chemically enrich earlier than the outskirts), inevitably results in the emergence of a negative metallicity gradient \citep{Boardman}. This scenario also supports efficient metal dilution, consistent with a closed galaxy evolution system \citep{Tissera_2022}. Conversely, positive radial $Z_{\rm gas}$ gradients can emerge in models where galaxies form and evolve within dense environments. In such environments, gas circulates dynamically through, around, and within galaxies \citep{Somerville}, potentially inverting the typical negative gradient. 

The presence of this dual behavior in disc metallicity gradients indicates that these physical ionised properties are not completely led by a single  mechanism in galaxy evolution. A negative gradient can be driven primarily by the inside-out formation of galaxies and the continuous accretion of metal-poor gas from the cosmic web, which dilutes metals more effectively in the outer regions. This is further supported by models such as those of  \cite{Tissera_2022} concerning efficient metal dilution in galaxy evolution. Conversely, the presence of a positive gradient hints towards the effect of environmental influences and gas movements inside galactic halos \citep{Somerville}. Therefore, the observed dual behavior is a strong indicator that galaxy evolution is a nuanced process, shaped by both internal evolutionary track and external environmental interactions.

In conclusion, our findings point to a limited role of Seyfert-driven outflows in shaping the metallicity redistribution on the galaxy. This may show that the outflows are not effective in displacing enriched material over large spatial scales. To investigate the role of outflows in redistributing metal contents, higher-resolution observations are essential to detect outflows with more details on its physical properties such as the total mass of the ionized gas, peak mass outflow rate, and total momentum. Such data would enable a more fundamental analysis of whether metal enrichment in the circumgalactic or ISM is primarily driven by outflow processes, or whether environmental factors and secular evolution play a more critical impacts on the distribution of metal content within galaxies.

\begin{acknowledgements}
We express our gratitude to the referee for improving this paper through their valuable suggestions. A.A. thanks Alessandro Marconi, Giacomo Venturi, Matilde Mingozzi, and  Giovanni Cresci for helpful discussions. Co-funded by the European Union (MSCA Doctoral Network EDUCADO, GA 101119830 and Widening Participation, ExGal-Twin, GA 101158446). JHK acknowledges grant PID2022-136505NB-I00 funded by MCIN/AEI/10.13039/501100011033
and EU, ERDF.
\end{acknowledgements}

%\clearpage
\FloatBarrier
\bibliographystyle{aa}
\bibliography{aanda}

\clearpage

\section{Appendix A: the $Z_{\rm gas}$ histogram}
\label{sec:appB}
\begin{figure*}[t]
  \centering
  \begin{tabular}{cc}
    \includegraphics[width=0.45\textwidth]{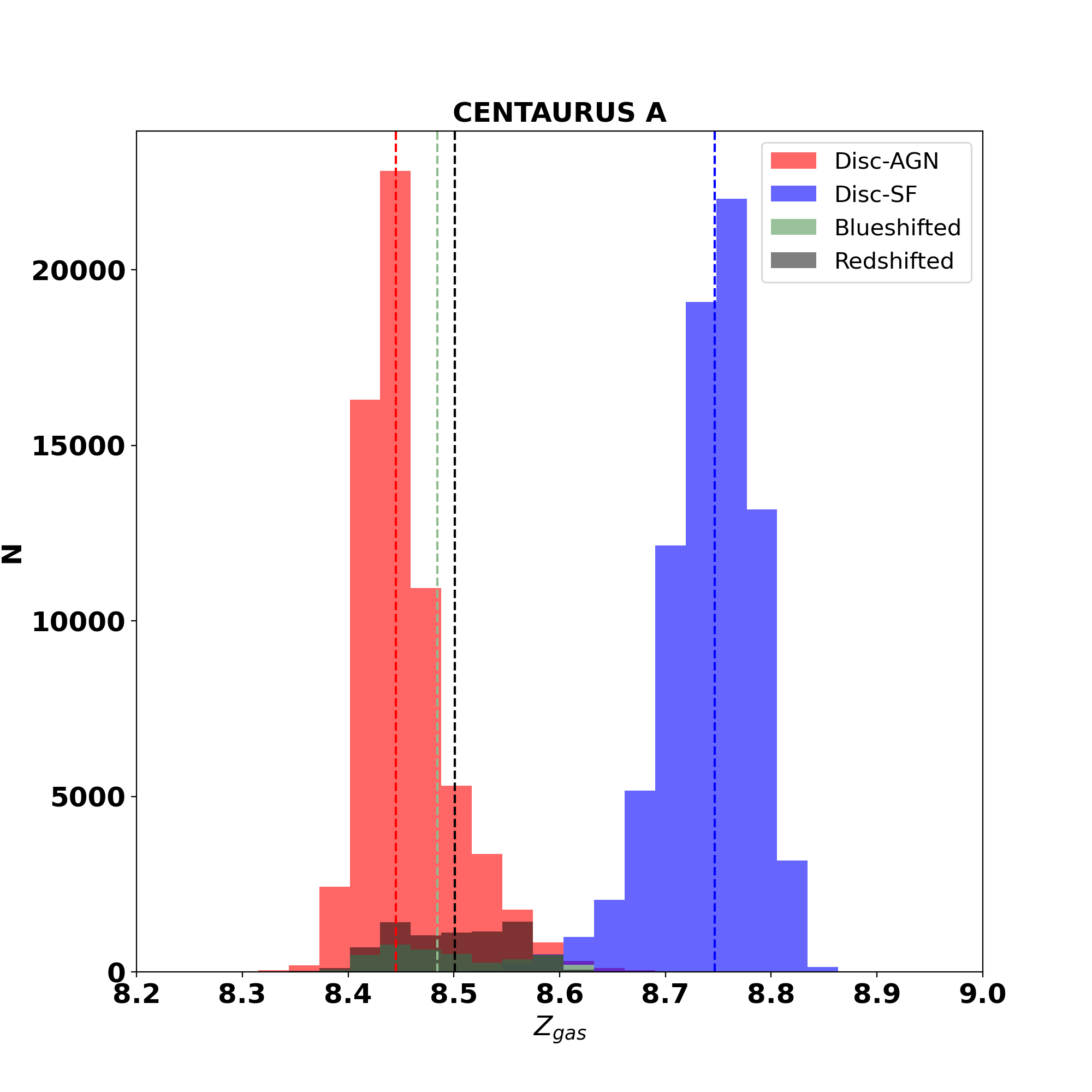} & \includegraphics[width=0.45\textwidth]{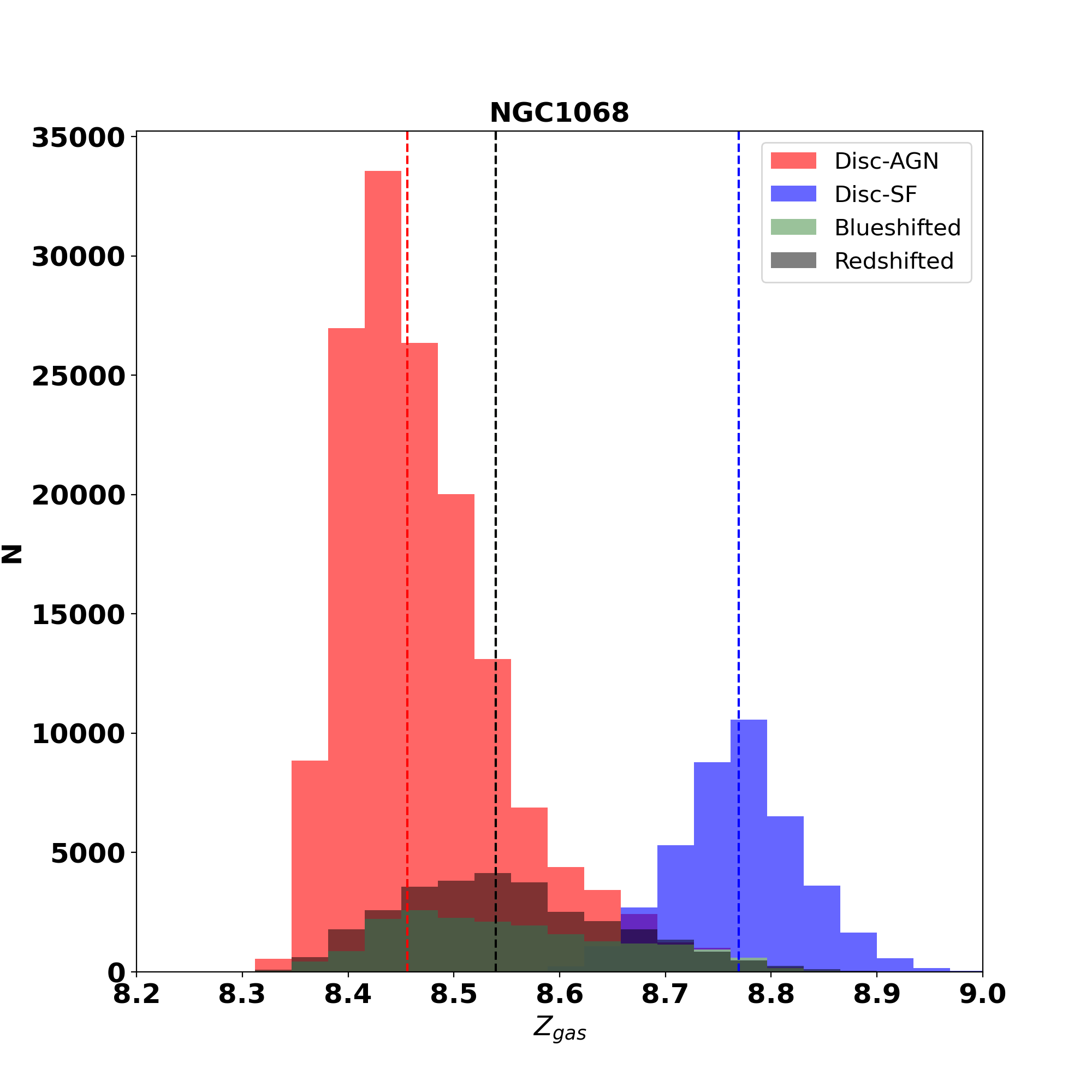} \\
    \includegraphics[width=0.45\textwidth]{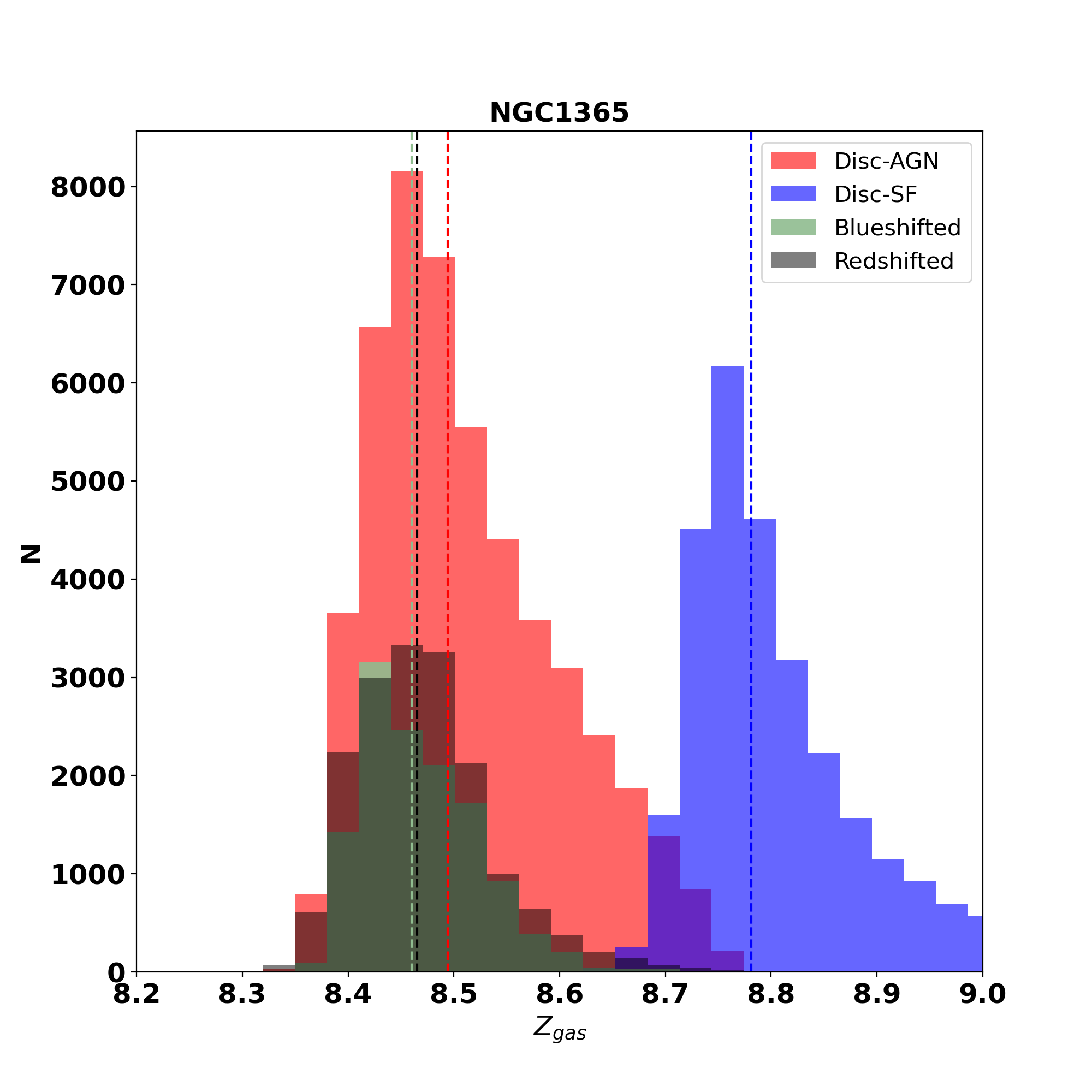} & \includegraphics[width=0.45\textwidth]{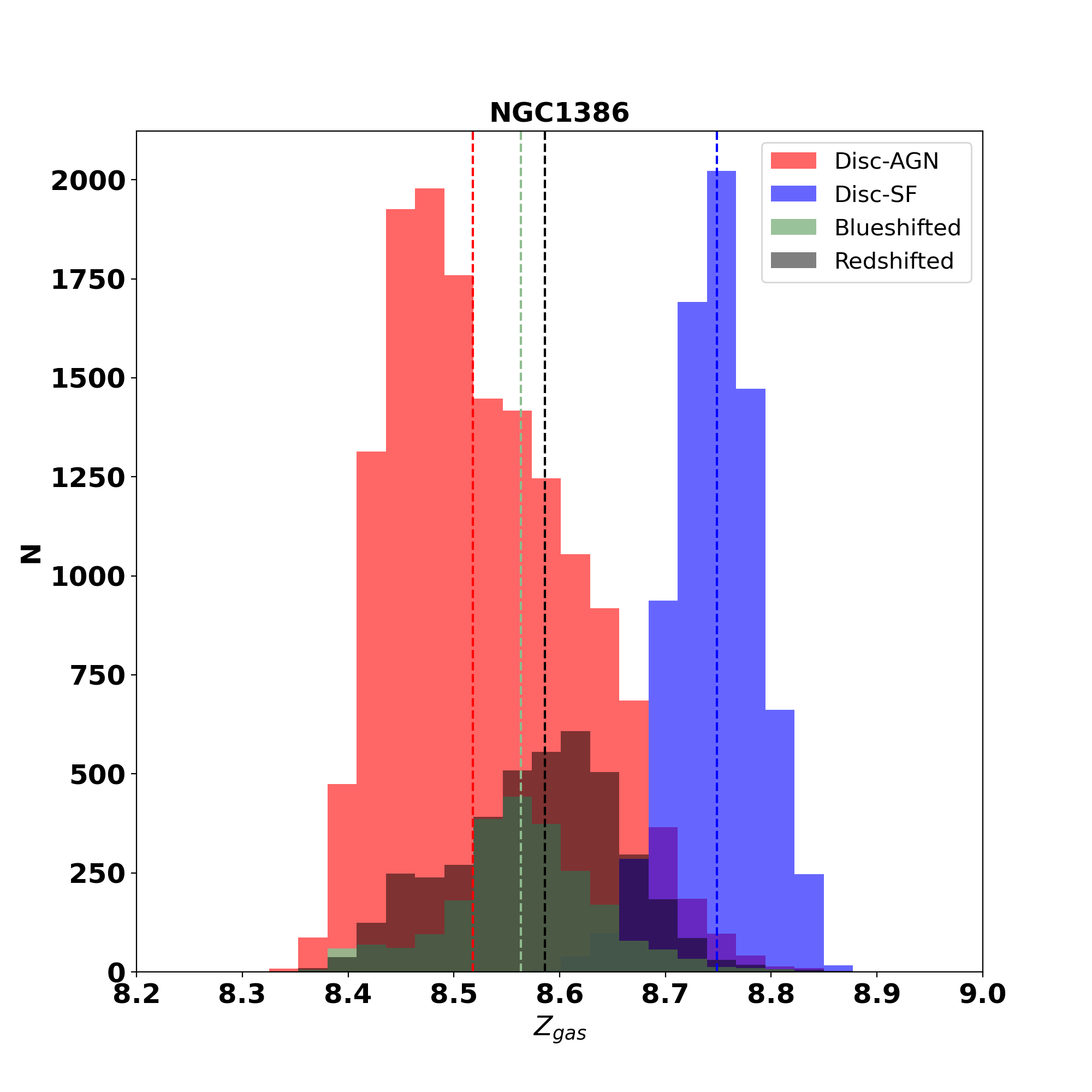} \\
    \includegraphics[width=0.45\textwidth]{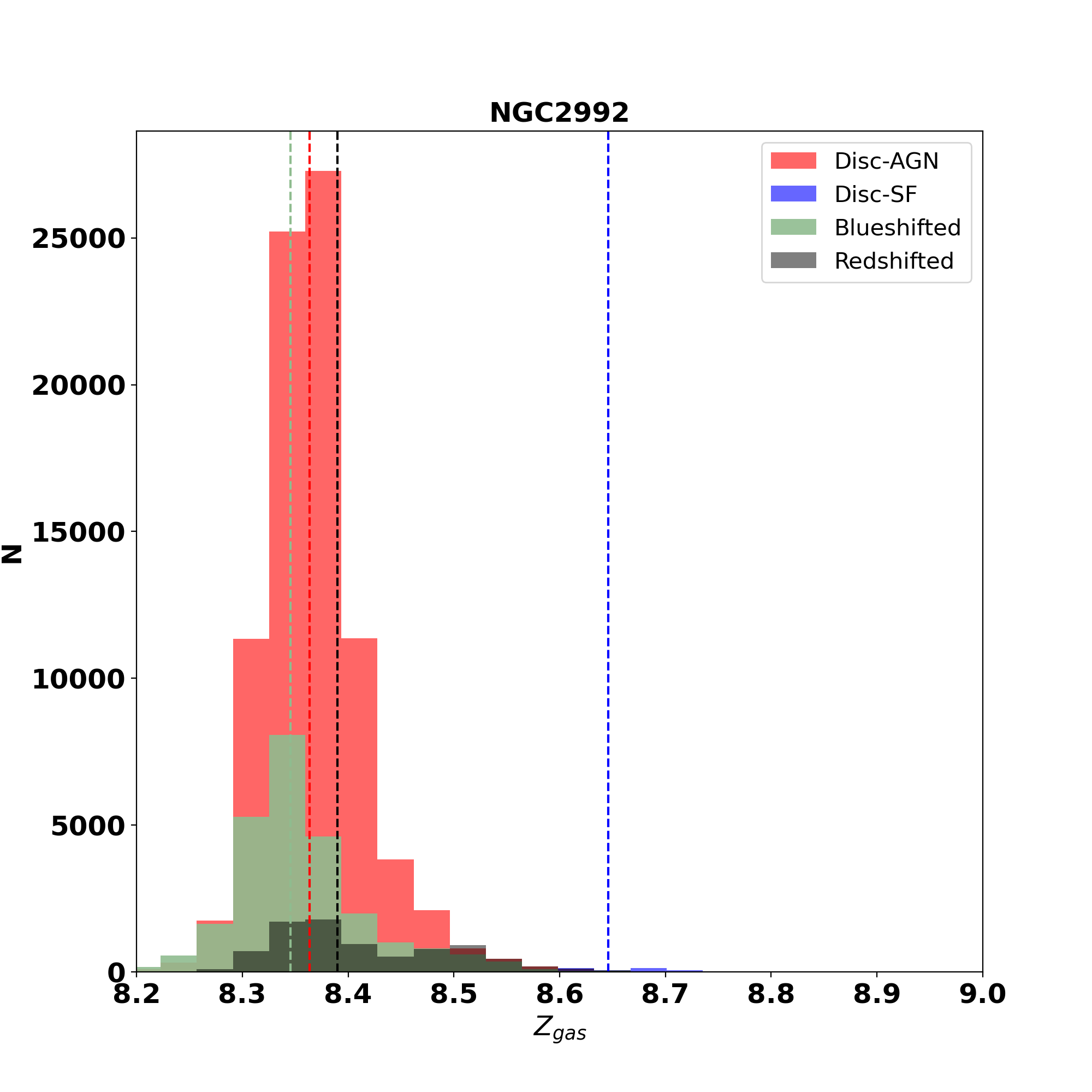} & \includegraphics[width=0.45\textwidth]{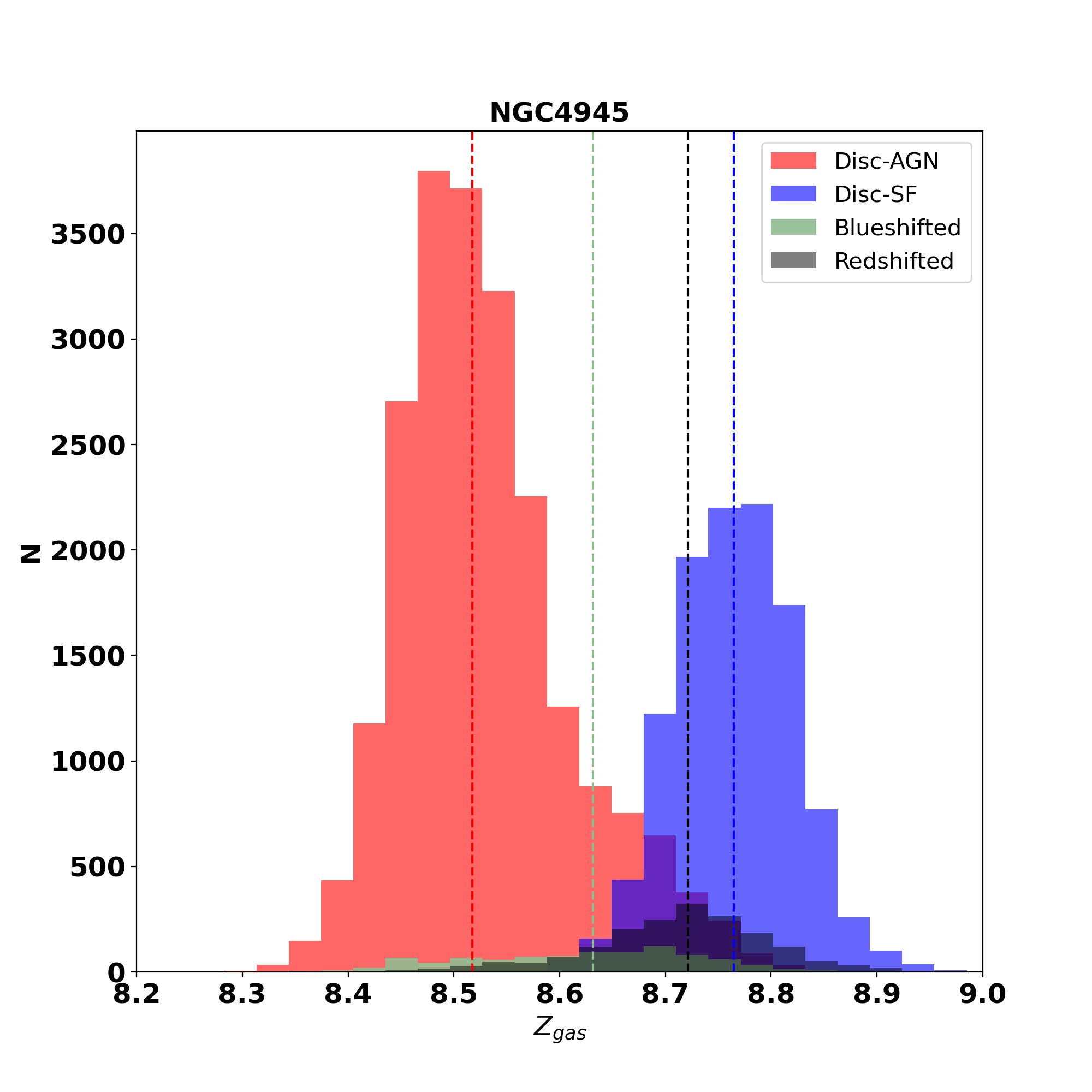} \\

   \end{tabular}
   \caption{Distribution of $Z_{\rm gas}$ in CENTAURUS A, NGC~1068, NGC~1365, NGC~1386, NGC~ 2992, and NGC~4945. The AGN parts are shown in disc(red), redshifted (black) and blueshifted (green) outflow components while the SF-regions in disk are plotted in blue color. The vertical dashed lines show the median value of $Z_{\rm gas}$  for each of the components in the same color.}
   \label{fig:hists_B1}
\end{figure*}

\begin{figure*}[t]
  \centering
  \begin{tabular}{cc}
  \includegraphics[width=0.45\textwidth]{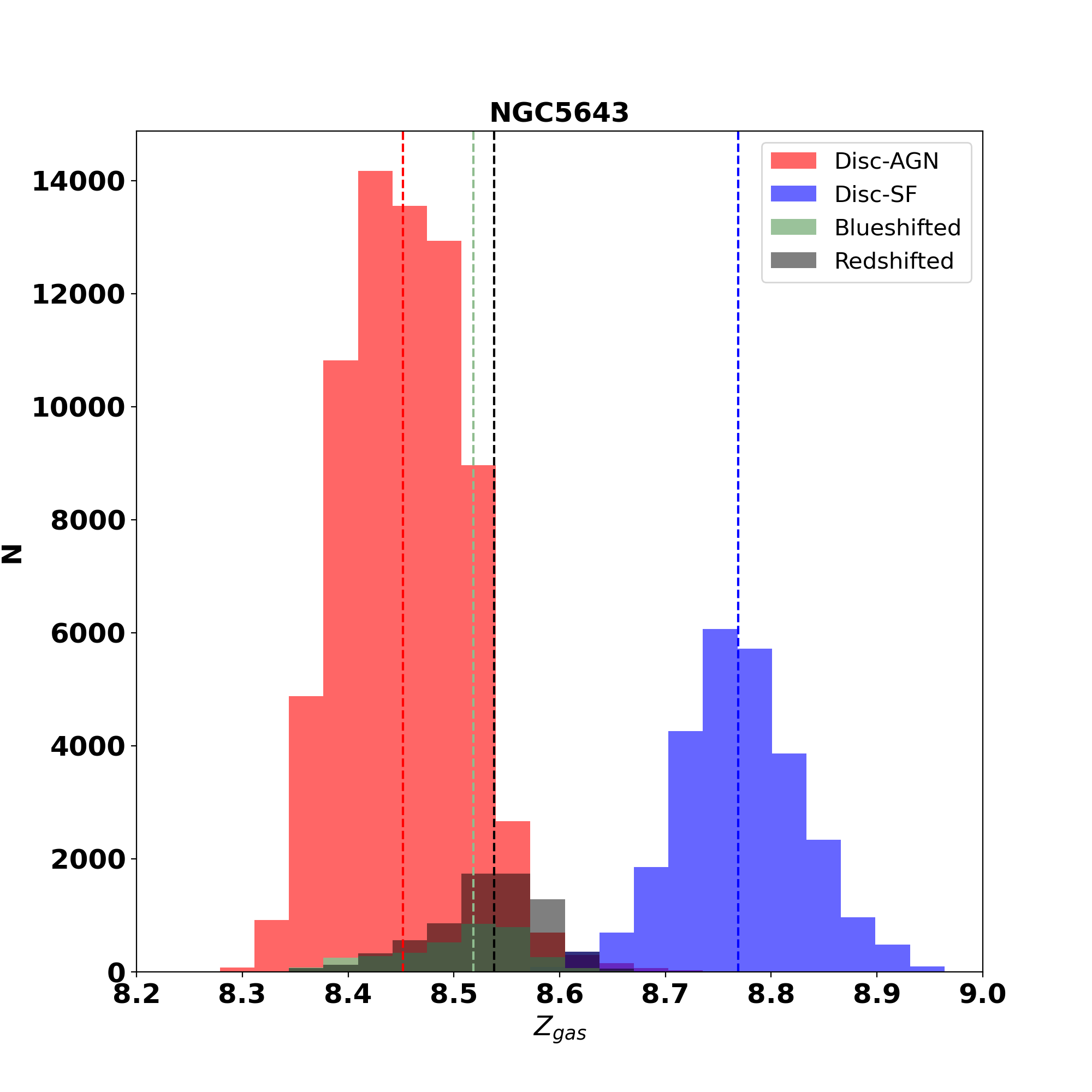} \\
  \end{tabular}
  \caption{The same as the Fig \ref{fig:hists_B1} but for NGC~5643.}
  \label{fig:hists_B2}
\end{figure*}

The distribution of $Z_{\rm gas}$ in CENTAURUS A, NGC~1068, NGC~1365, NGC~1386, NGC~2992, NGC~4945, and NGC~5643 are shown in Fig~\ref{fig:hists_B1}, and \ref{fig:hists_B2} as described in Sec. \ref{sec:results_II}. We consider the \citet{Storchi-Bergmann_1998} and \citet{pettini} calibration relations to compute $Z_{\rm gas}$ for AGNs and SF regions, respectively.

\end{document}